\documentclass[a4paper,UKenglish,cleveref, autoref, thm-restate]{lipics-v2021}

\hideLIPIcs

\usepackage{mathtools} 
\usepackage{amssymb}
\usepackage{mathrsfs}
\usepackage{anyfontsize}
\usepackage{complexity}

\usepackage[linesnumbered,lined,ruled,noend,vlined]{algorithm2e}

\let\oldReturn\Return 
\let\Return\undefined 
\usepackage{ascmac}
\usepackage{fancybox}
\let\Return\oldReturn

\usepackage{tcolorbox}
\usepackage{fbox}

\usepackage{tikz}
\usetikzlibrary{math,calc,arrows.meta,bending,intersections,positioning}

\usepackage[mathlines]{lineno}
\linenumbers
\usepackage[titletoc]{appendix}

\newcommand{\rev}[1]{#1}
\newcommand{\revi}[1]{#1}

\crefname{figure}{Figure}{Figures}
\crefname{theorem}{Theorem}{Theorems}
\crefname{lemma}{Lemma}{Lemmas}
\crefname{claim}{Claim}{Claims}
\crefname{observation}{Observation}{Observations}
\crefname{redrule}{Reduction Rule}{Reduction Rules}
\crefname{branching}{Branching}{Branchings}
\crefname{enumi}{claim}{claims}
\Crefname{enumi}{Claim}{Claims}

\newtheorem{redrule}{Reduction}
\newtheorem{branching}{Branching}

\newcommand{\prb}[1]{\textnormal{\scshape #1}}

\newcommand{\ttyes}{\texttt{YES}}
\newcommand{\ttno}{\texttt{NO}}

\newcommand{\excess}[1]{\textsf{ex}(#1)}

\newcommand{\contcomp}[1]{\textsf{cc}(#1)}
\newcommand{\dist}[3]{\textsf{dist}_{#1}({#2},{#3})}
\newcommand{\singletons}[1]{S_{#1}}
\DeclareMathOperator{\symdif}{\triangle}

\newcommand{\rmpi}{\mathrm{\Pi}}
\newcommand{\Lclass}{\mathscr{L}}

\newcommand{\delprb}{$\mathcal{F}$\prb{-Deletion Set}}
\newcommand{\fvsprb}{\prb{Feedback Vertex Set}}
\newcommand{\confgraph}[2]{G_{#1,#2}}

\newenvironment{listing}[1]{%
  \begin{list}{*}{%
    \settowidth{\labelwidth}{#1}%
    \setlength{\leftmargin}{\labelwidth}%
    \advance \leftmargin by 12pt
    \setlength{\itemsep}{0pt}%
    \setlength{\parsep}{0pt}%
    \setlength{\topsep}{0pt}%
    \setlength{\parskip}{0pt}%
  }%
}{%
  \end{list}%
}

\SetKwFunction{ComputingProductList}{ComputeList}
\SetKwFunction{Search}{Search}
\SetKwFunction{InSolution}{InSolution}
\SetKwFunction{NotInSolution}{NotInSolution}

\newcommand{\Qrank}{\textsf{rank}}
\newcommand{\lab}{\textsf{label}}

\title{\texorpdfstring{$k$}{k}-Product Lists for (Generalized) Feedback Vertex Sets: A Tool for Designing FPT Algorithms} 

\titlerunning{\texorpdfstring{$k$}{k}-Product Lists for (Generalized) Feedback Vertex Sets} 

\author{Rin Saito}{Graduate School of Information Sciences, Tohoku University, Sendai, Japan}{rin.saito@dc.tohoku.ac.jp}{https://orcid.org/0000-0002-3953-4339}{This work was supported by JSPS KAKENHI Grant Numbers JP26KJ0550 and JP24H00686.}

\author{Yuma Tamura}{Graduate School of Information Sciences, Tohoku University, Sendai, Japan}{tamura@tohoku.ac.jp}{https://orcid.org/0009-0001-5479-7006}{This work was supported by JSPS KAKENHI Grant Numbers JP25K21148 and JP24H00686.}

\authorrunning{R. Saito and Y. Tamura} 

\Copyright{Rin Saito and Yuma Tamura} 



\category{} 

\relatedversion{} 

\nolinenumbers 

\EventEditors{Tuukka Korhonen and Michael Lampis}
\EventNoEds{2}
\EventLongTitle{21st International Symposium on Parameterized and Exact Computation (IPEC 2026)}
\EventShortTitle{IPEC 2026}
\EventAcronym{IPEC}
\EventYear{2026}
\EventDate{September 2--4, 2026}
\EventLocation{L’Aquila, Italy}
\EventLogo{}
\SeriesVolume{398}
\ArticleNo{25}

\begin{document}

\maketitle

\begin{abstract}
Enumerating all inclusion-minimal solutions is a powerful approach to designing fixed-parameter tractable (FPT) algorithms. 
However, for the \textsc{Feedback Vertex Set} problem, this approach faces a fundamental obstacle, as the number of minimal feedback vertex sets is not, in general, bounded by any function of the solution size~$k$. 
To overcome this limitation, we introduce a data structure called a \emph{$k$-product list}, which implicitly encodes all minimal solutions while admitting non-minimal ones to maintain efficiency. 
We present FPT algorithms for computing such lists not only for the \textsc{Feedback Vertex Set} problem but also for its natural generalizations, the \textsc{$\ell$-Forest Deletion Set} problem and the \textsc{$r$-Pseudoforest Deletion Set} problem.
Our framework yields simple FPT algorithms for a wide range of variants, providing the first such results in several cases and demonstrating the broad applicability of our approach.
\end{abstract}

\clearpage
{\tableofcontents}
\clearpage

\section{Introduction}

Given a graph $G = (V, E)$, a \emph{feedback vertex set} is a vertex subset $F$ of $G$ such that the subgraph of $G$ induced by $V \setminus F$ is a forest.
The \prb{Feedback Vertex Set} problem asks, given a graph $G$ and a positive integer $k$, whether $G$ has a feedback vertex set of size at most $k$.
This problem is NP-complete~\cite{FVSkarp:Karp72} and has been studied extensively in parameterized complexity, particularly with respect to the design of faster fixed-parameter tractable (FPT) algorithms~\cite{FVSFPT:IwataK21,FVSFPT:KociumakaP14,FVSFPT:LiN22} and the improvement of polynomial kernels~\cite{FVSkernel:Iwata17,FVSkernel:Thomasse10}.

For combinatorial problems, it is natural to consider variants obtained by imposing additional structural constraints on feasible solutions.
For example, in the \prb{Independent Feedback Vertex Set} problem (resp.~\prb{Connected Feedback Vertex Set}), the task is to find a feedback vertex set $F$ such that the subgraph induced by $F$ is edgeless (resp.~connected).
Beyond such structural variants, several recent lines of research are motivated by practical considerations and study combinatorial problems within broader problem frameworks, including \emph{combinatorial reconfiguration}~\cite{reconf:ItoDHPSUU11} and \emph{diverse solutions}~\cite{diverse:BasteJMPR19}.

One standard approach to designing FPT algorithms parameterized by the solution size $k$ for such variants is to enumerate all inclusion-minimal (feasible) solutions of size at most $k$.
If this enumeration can be carried out in time $f(k) \cdot n^{O(1)}$, where $f$ is some function, then the number of such minimal solutions is bounded by $f(k)\cdot n^{O(1)}$, which often enables efficient algorithms for the corresponding variants.
This phenomenon is well known for \prb{Vertex Cover} and, more generally, for \prb{$d$-Hitting Set} with fixed $d$, where simple branching algorithms yield an FPT-time enumeration of all minimal solutions~\cite[Section~1.3]{book:FlumG06}.
Although such algorithms are sometimes considered naive, they frequently serve as a starting point for developing faster or more refined algorithms.

We now return to the \prb{Feedback Vertex Set} problem.
Unlike \prb{Vertex Cover}, minimal feedback vertex sets cannot, in general, be enumerated in FPT time parameterized by the solution size $k$.
To see this, consider a graph consisting of $k$ disjoint cycles, each of length $\lceil n/k \rceil$.
Such a graph contains $\Omega((n/k)^k)$ distinct minimal feedback vertex sets, implying that the output size is not bounded by any function of $k$ alone.
Consequently, FPT-time enumeration of all minimal feedback vertex sets is impossible.
\footnote{This output-size argument concerns the total time required to list the entire family; it does not rule out polynomial-delay enumeration, which is outside the scope of this paper.}

In the course of designing faster FPT algorithms for \prb{Feedback Vertex Set}, Guo et al.~\cite{CompactRep:GuoGHNW06} introduced the notion of a \emph{list of compact representations} as a by-product.
Intuitively, a list of compact representations is a data structure of FPT size that implicitly encodes a family of minimal feedback vertex sets.

Formally, a \emph{compact representation} of a graph $G$ is a family $\mathcal{C}$ of pairwise disjoint vertex subsets of $G$ such that selecting exactly one vertex from each set in $\mathcal{C}$ always yields a minimal feedback vertex set of $G$.
A minimal feedback vertex set $S$ is said to be \emph{represented} by $\mathcal{C}$ if $|S \cap C| = 1$ for every $C \in \mathcal{C}$.
For a positive integer $k$, a compact representation $\mathcal{C}$ is called \emph{$k$-compact} if $|\mathcal{C}| \leq k$.
A collection $\mathcal{L}$ of $k$-compact representations is called a \emph{$k$-compact list of feedback vertex sets} if every minimal feedback vertex set of $G$ \revi{of size at most $k$} is represented by at least one $\mathcal{C} \in \mathcal{L}$.
Guo et al.~\cite{CompactRep:GuoGHNW06} showed that a $k$-compact list of FPT size can be computed in FPT time parameterized by $k$.
This notion has since become a useful tool in the design of FPT algorithms for several variants of \prb{Feedback Vertex Set}~\cite{diverse:Kumabe25,connected:MisraPRSS12,SolDiscv:SaitoSSST25,hereditary:TamuraI020}.

\subsection*{Our Contribution}
The notion of a $k$-compact list provides a powerful implicit representation of all minimal feedback vertex sets.
A crucial limitation, however, is that each compact representation is required to generate only minimal solutions.
This restriction, which arises from the definition of a $k$-compact representation itself, is theoretically natural but often too strong for practical algorithmic applications.
In many such settings, it suffices to include all minimal solutions, even if the representation also produces non-minimal ones.

Motivated by this observation, we introduce a relaxed notion of a compact representation.
Specifically, we drop the requirement that selecting one vertex from each set of the compact representation yields a \emph{minimal} feedback vertex set, and instead allow for a \emph{non-minimal} feedback vertex set.
Crucially, we retain the original requirement of a $k$-compact list: every minimal feedback vertex set must still be represented by at least one element of the list.

We call the resulting structure a \emph{$k$-product representation}, and the corresponding collection a \emph{$k$-product list} of feedback vertex sets.
In \Cref{sec:application}, we demonstrate that this relaxed notion still retains substantial algorithmic usefulness.
In particular, we show that this notion suffices to enable the design of FPT algorithms for a wide range of variants of \prb{Feedback Vertex Set}, including weighted, connected, non-partisan, hereditary-property, conflict-free, simultaneous, reconfiguration, and diverse variants.

Our contribution is not centered on \prb{Feedback Vertex Set} itself, but rather on the development of a unifying data-structural framework for efficiently encoding all minimal solutions.
To demonstrate the generality of our approach, we consider two extensions of feedback vertex sets, namely \emph{$\ell$-forest deletion sets} and \emph{$r$-pseudoforest deletion sets}, where $\ell$ and $r$ are nonnegative integers.
(See~\Cref{sec:Preliminaries} for formal definitions; when $\ell = r = 0$, these notions coincide with feedback vertex sets.)
It is known that deciding whether a given graph admits such a deletion set of size at most $k$ is fixed-parameter tractable when parameterized by $k+\ell$ and $k+r$, respectively~\cite{lforest:RaiS18,rpseudoforest:ShengG23}.
We show that $k$-product lists for these generalizations can also be constructed in time $O^*(64^\ell \cdot 17^k)$\footnote{We use $O^*$ notation to suppress polynomial factors in the input size.} and $O^*(c_r^k)$ for $\ell \geq 0$ and $r \geq 1$, respectively, where $c_r$ is a constant depending only on $r$.
As a corollary, we obtain an algorithm for computing a $k$-product list of feedback vertex sets in time $O^*(17^k)$.
This bound has a smaller exponential base than the previously known $O^*(23.1^k)$ bound of Guo et al.~\cite{CompactRep:GuoGHNW06,connected:MisraPRSS12} for constructing a $k$-compact list for feedback vertex sets.
A refined analysis of our algorithms using the measure-and-conquer technique~\cite{FominGK09} further improves the running time for constructing a $k$-product list for $\ell$-forest deletion sets to $O^*(14.4^\ell \cdot 13^k)$, yielding a running time of $O^*(13^k)$ for feedback vertex sets.

As a consequence, our framework yields FPT algorithms for a wide range of variants of $\ell$-forest deletion sets and $r$-pseudoforest deletion sets, including the weighted, nonpartisan, hereditary-constraint, reconfiguration, connected, conflict-free, simultaneous, and diverse variants (see \Cref{sec:application} for definitions).
For the connected and diverse variants, existing algorithms based on $k$-compact lists extend directly to our setting.
For the simultaneous, weighted, nonpartisan, reconfiguration, and conflict-free variants, we develop new algorithms based on $k$-product lists.
For the hereditary-constraint variant, we build upon the approach of Tamura et al.~\cite{hereditary:TamuraI020} for the corresponding variant of \prb{Feedback Vertex Set}, and extend it using structural properties specific to $\ell$-forest deletion sets and $r$-pseudoforest deletion sets.
While all these variants are known to be fixed-parameter tractable for \prb{Feedback Vertex Set}, fixed-parameter tractability for the corresponding variants of $\ell$-forest deletion sets and $r$-pseudoforest deletion sets was previously unknown.
Moreover, several of our algorithms run in single-exponential time in $k$ when $\ell$ or $r$ is treated as a constant.

\subsection*{Technical Overview}
We first briefly explain the framework by Guo et al.~\cite{CompactRep:GuoGHNW06} for computing a compact list for the \prb{Feedback Vertex Set} problem.
Their algorithm begins by computing a feedback vertex set $S$ of size at most $k$ using an existing algorithm.
Subsequently, for every subset $S' \subseteq S$, it constructs a compact list representing the minimal feedback vertex sets whose intersection with $S$ is exactly $S'$.
Fix such a subset $S'$ and consider the graph $G' \coloneq G - S'$.
Any path $P$ of length at least two whose internal vertices all have degree two is contracted into a single set representing the component $V(P)$ in the compact representation.
After further deleting vertices of degree at most~$1$, the number of vertices in $G' - (S \setminus S')$ can be bounded by a function of $k$.
Finally, the algorithm constructs the compact representation by exhaustively selecting valid vertices from $G' - (S \setminus S')$.

The correctness of the path contraction follows because any minimal feedback vertex set intersects $P$ in at most one vertex, implying that contracting $P$ preserves all minimal solutions.
While the approach of Guo et al.\ constructs a compact representation via long-path contractions, the exhaustive-search step remains the main computational bottleneck.

Fixing a partial solution $S'$ and exploring the remaining graph $G - S'$ is a central idea in \emph{iterative compression}~\cite{textbook:CyganFKLMPPS15,iterative:ReedSV04}, in particular in its disjoint variant.
In state-of-the-art algorithms for \prb{Feedback Vertex Set} and its generalizations, this disjoint variant is typically solved using carefully designed branching rules rather than exhaustive enumeration.
Motivated by this, we replace the exhaustive phase in the algorithm of Guo et al.\ with a branching algorithm inspired by methods for \prb{Disjoint Feedback Vertex Set}.

A key technical challenge is to preserve the information of all minimal solutions.
Standard branching rules for \prb{Disjoint Feedback Vertex Set} are designed to compute a single optimal solution and may safely discard branches that do not lead to such a solution.
However, in our setting, such pruning is not permissible, as it may eliminate branches corresponding to other minimal solutions.

To address this issue, we modify the reduction and branching rules so that all minimal solutions are preserved.
Although these modifications may introduce some non-minimal solutions into the resulting representation, they guarantee that all minimal solutions are retained.
Finally, by analyzing the resulting search tree with an appropriate measure, we show that the overall algorithm still runs in FPT time.
Moreover, by employing appropriate branching rules, this strategy can be extended beyond \prb{Feedback Vertex Set} to its generalizations such as \prb{$\ell$-Forest Deletion Set} and \prb{$r$-Pseudoforest Deletion Set}.

\medskip
\noindent
\textbf{Related work.}
For the \prb{Feedback Vertex Set} problem, the fastest known deterministic parameterized algorithm, due to Iwata and Kobayashi~\cite{FVSFPT:IwataK21}, runs in time $O^*(3.460^k)$, 
whereas the fastest randomized algorithm, due to Li and Nederlof~\cite{FVSFPT:LiN22}, runs in time $O^*(2.7^k)$.

The notion of a \emph{$k$-compact list} was introduced by Guo et al.~\cite{CompactRep:GuoGHNW06} as a data structure for representing all minimal solutions.
Later, Misra et al.~\cite{connected:MisraPRSS12} improved the upper bound on the size of a $k$-compact list to $O^*(23.1^k)$, strengthening the original bound of Guo et al.
Since then, $k$-compact lists have been used to design FPT algorithms for various problems~\cite{diverse:Kumabe25,connected:MisraPRSS12,SolDiscv:SaitoSSST25,hereditary:TamuraI020}.

The \prb{$\ell$-Forest Deletion Set} problem, also known as \prb{Almost Forest Deletion}, was introduced by Rai and Saurabh~\cite{lforest:RaiS18}, who gave an $O^*(4.0024^k \cdot 5.0018^{\ell})$-time algorithm and an $O(k\ell(k+\ell))$-vertex kernel.
Lin et al.~\cite{lforest:LinFWCFL18} later improved the running time to $O^*(5^k \cdot 4^{\ell})$.
Recently, Gowda et al.~\cite{lforest:GowdaLPPS24} obtained a randomized algorithm with running time
${\mathcal O}^*\!\left(\min\{2.85^k \cdot 8.54^{\ell},\; 2.7^k \cdot 36.61^{\ell},\; 3^k \cdot 1.78^{\ell}\}\right)$.

The \prb{$r$-Pseudoforest Deletion Set} problem was introduced by Philip et al.~\cite{rpseudoforest:PhilipRS18}, who showed that it is solvable in $O^*(c_r^k)$ (where $c_r$ is a constant depending only on $r$) and admits a kernel of size $f(r)\cdot k^2$ for some function $f$.
Sheng and Gutin~\cite{rpseudoforest:ShengG23} subsequently obtained an $O^*((1 + (2r + 3)^{r + 2})^{k + 1})$-time algorithm, which was improved by Tsur~\cite{rpseudoforest:Tsur25} to $O^*((8r+1)^k)$.
For $r=1$, the problem is known as \prb{Pseudoforest Deletion Set}. 
Bodlaender et al.~\cite{pseudoforest:BodlaenderOO18} gave an $O^*(3^k)$-time algorithm, and Gowda et al.~\cite{lforest:GowdaLPPS24} later obtained a randomized $O^*(2.85^k)$-time algorithm.


\medskip
\noindent
\textbf{Organization.}
After the preliminaries (\Cref{sec:Preliminaries}), we introduce our algorithmic framework in \Cref{sec:framework}.
We then apply it to $\ell$-forest deletion sets in \Cref{sec:lforest} and to $r$-pseudoforest deletion sets in \Cref{sec:pseudoforest}.
In \Cref{sec:application}, we further demonstrate the usefulness of $k$-product lists by giving FPT algorithms for some variants.
Finally, we conclude with some remarks in \Cref{sec:remarks}.

\medskip
\noindent
\textbf{Declaration of generative AI use.}
The refined running-time analysis of \Cref{thm:lforest_comp_rep} was derived by OpenAI's GPT-6 based on pseudocode provided by the authors. 
In particular, GPT-6 derived the optimized parameters for the measure-and-conquer analysis and the corresponding running-time bounds. 
The authors reviewed and verified the resulting analysis.

\section{Preliminaries}
\label{sec:Preliminaries}
For a positive integer~$n$, we write $[n] \coloneqq \{1,2,  \ldots, n\}$.

Let $G$ be a multigraph, that is, $G$ may contain parallel edges.
Let $V(G)$ and $E(G)$ denote its vertex set and edge set, respectively.
For a vertex subset $S$ of $G$, let $G[S]$ denote the subgraph of $G$ induced by $S$, and let $G-S$ be the graph obtained by deleting all vertices in $S$ together with their incident edges; for simplicity, we write $G-v$ instead of $G-\{v\}$ for $v \in V(G)$.
For an edge subset $F$ of $G$, let $G-F$ denote the graph obtained by deleting all edges in $F$.
A vertex subset $U$ of $G$ is \emph{connected} if $G[U]$ is connected.
A \emph{connected component} of $G$ is a maximal vertex subset $C \subseteq V(G)$ such that $G[C]$ is connected, and the number of connected components of $G$ is denoted by $\contcomp{G}$.

For a vertex $v \in V(G)$, the \emph{open neighborhood} of $v$ is the set $N(v)$ of vertices adjacent to $v$, and the \emph{closed neighborhood} of $v$ is $N[v] \coloneqq N(v) \cup \{v\}$.
For $X \subseteq V(G)$, let $N_X(v) \coloneqq N(v) \cap X$.
\revi{
For $v \in V(G)$, let $\deg_G(v)$ denote its \emph{degree}.
For $S \subseteq V(G)$, let $\deg_S(v)$ denote the number of incidences at $v$ whose other endpoint belongs to $S$.
Thus, a loop at $v$ contributes two to $\deg_S(v)$ when $v \in S$.
}
\revi{For $U \subseteq V(G)$, define $\deg_{S}(U) \coloneq \sum_{v \in U} \deg_{S}(v)$.}
For an edge $e = uv \in E(G)$, let $G/e$ denote the graph obtained by \emph{contracting} $e$, that is, by identifying $u$ and $v$. 
For vertices $u, v \in V(G)$, the \emph{distance} $\dist{G}{u}{v}$ is the number of edges in a shortest $uv$-path in $G$, and the \emph{diameter} of $G$ is $\max_{u,v \in V(G)} \dist{G}{u}{v}$.

A \emph{forest} is a graph containing no cycles.
For a nonnegative integer $\ell$, an \emph{$\ell$-forest} is a graph $G$ that can be made a forest by deleting at most $\ell$ edges; that is, there exists $F \subseteq E(G)$ with $|F| \le \ell$ such that $G-F$ is a forest.
For a nonnegative integer $r$, an \emph{$r$-pseudoforest} is a graph whose every connected component is an $r$-forest.
In particular, $0$-forests and $0$-pseudoforests coincide with forests.
For a connected component $C$ of $G$, the \emph{excess} of $C$ is defined as
$\excess{C} \coloneq |E(G[C])| - |C| + 1$.
The \emph{excess} of $G$ is
$\excess{G} \coloneq \sum_{C \in \mathcal{C}} \excess{C}$,
where $\mathcal{C}$ is the set of connected components of $G$; equivalently, it is the minimum number of edges whose removal makes $G$ a forest.
Note that $G$ is an $\ell$-forest if and only if $\excess{G} \le \ell$.

Let $\mathcal{F}$ be a graph class.
A vertex subset $S$ of a graph $G$ is an \emph{$\mathcal{F}$-deletion set} if $G - S \in \mathcal{F}$.
It is \emph{minimal} if no proper subset of $S$ is an $\mathcal{F}$-deletion set.
Given a graph $G$ and an integer $k$, the \delprb\ problem asks whether there exists an $\mathcal{F}$-deletion set of size at most $k$.
When $\mathcal{F}$ is the class of forests, $\ell$-forests, or $r$-pseudoforests, we refer to $\mathcal{F}$-deletion sets as \emph{feedback vertex sets}, \emph{$\ell$-forest deletion sets}, and \emph{$r$-pseudoforest deletion sets}, respectively.

A graph class $\mathcal{F}$ is \emph{closed under taking subgraphs} if every subgraph of a graph in $\mathcal{F}$ also belongs to $\mathcal{F}$.
If $\mathcal{F}$ has this property, then for any $\mathcal{F}$-deletion set $S$ of $G$ and any vertex $v \in V(G)$, the set $S \cup \{v\}$ is also an $\mathcal{F}$-deletion set of $G$.

A graph $H$ is a \emph{minor} of $G$ if there exists a \emph{minor model} $\phi$ of $H$ in $G$, that is, a mapping $\phi$ that assigns to each vertex $v \in V(H)$ a connected subgraph $\phi(v)$ of $G$, and to each edge $e \in E(H)$ an edge $\phi(e) \in E(G)$ such that
\begin{enumerate}
    \item for all distinct $u, v \in V(H)$, we have $V(\phi(u)) \cap V(\phi(v)) = \emptyset$, and
    \item for every edge $e = uv \in E(H)$, the edge $\phi(e) = u'v' \in E(G)$ satisfies $u' \in V(\phi(u))$ and $v' \in V(\phi(v))$.
\end{enumerate}
The subgraph $\phi(v)$ for a vertex $v \in V(H)$ is called the \emph{branch set} of $v$.

A \emph{product representation} of $\mathcal{F}$-deletion sets of a graph $G$ is a family $\mathcal{C}$ of pairwise disjoint, nonempty subsets of $V(G)$ such that every set obtained by selecting exactly one vertex from each member of $\mathcal{C}$ is an $\mathcal{F}$-deletion set of $G$.
\revi{We say that $\mathcal{C}$ \emph{represents} a set $S\subseteq V(G)$ if $S\cap C\neq\emptyset$ for every $C\in\mathcal{C}$.}
If $\mathcal{F}$ is closed under taking subgraphs, then every set represented by a product representation is an $\mathcal{F}$-deletion set: it contains a deletion set obtained by selecting one vertex from each member of the representation.
A product representation $\mathcal{C}$ is a \emph{$k$-product representation} if $|\mathcal{C}| \le k$, and each $C \in \mathcal{C}$ is called a \emph{component}.
A collection $\mathscr{L}$ of $k$-product representations \emph{represents} an $\mathcal{F}$-deletion set $S$ if $S$ is represented by some $\mathcal{C} \in \mathscr{L}$.
We call $\mathscr{L}$ a \emph{$k$-product list for $\mathcal{F}$-deletion sets} of $G$ if it represents every minimal $\mathcal{F}$-deletion set of $G$ with size at most $k$.

\revi{Observe that if a product representation $\mathcal{C}$ represents a minimal $\mathcal{F}$-deletion set $S$, then $S$ contains exactly one vertex from each member of $\mathcal{C}$ and no other vertices.  
Indeed, selecting one vertex from each set $S\cap C$ yields an $\mathcal{F}$-deletion set contained in $S$, which must equal $S$ by minimality.}

\subsection{\texorpdfstring{$\ell$}{ell}-Forest Deletion Set}
Here an additional observation for $\ell$-forest deletion sets is provided. 

\begin{lemma}[{\cite[Lemma~3]{lforest:RaiS18}}]
\label{fds:subgraph}
If a vertex $v \in V(G)$ is not contained in any cycle of $G$, then
$\excess{G - \{v\}} = \excess{G}$.
Moreover, if there exists a cycle in $G$ that contains $v$, then
$\excess{G - \{v\}} \leq \excess{G} - 1$.
\end{lemma}

\subsection{$r$-Pseudoforest Deletion Set} 
Here an additional observation for $r$-pseudoforest deletion sets is provided. 

\begin{observation}[Observation~1 in~\cite{rpseudoforest:PhilipRS18}]
\label{obs:rpseudo_subgraph}
Let $G'$ be a subgraph of $G$.
If $G$ is an $r$-pseudoforest, then $G'$ is also an $r$-pseudoforest.
\end{observation}

\subsection{Branching}
We consider branching algorithms for decision problems, where each instance $I$ is equipped with a nonnegative real-valued measure $\mu(I)$.
A branching step replaces an instance $I$ with subinstances $I_1, I_2, \ldots, I_t$ such that $I$ is a $\ttyes$-instance if and only if at least one of these subinstances is a $\ttyes$-instance.
For the running-time analysis, we require that $\mu(I_j)<\mu(I)$ for every $j\in [t]$.

Let $\mu(I)=d$, and suppose a branching produces $t$ subinstances with measures $\mu(I_1)=d_1, \mu(I_2)=d_2, \ldots, \mu(I_t)=d_t$, respectively.
The tuple $(d-d_1, d-d_2, \ldots, d-d_t)$ is called the \emph{branching vector} of the branching.
The \emph{branching root} is the unique positive real solution to the equation $x^d-x^{d_1}-x^{d_2}-\cdots-x^{d_t}=0$.
We say that a branching is \emph{superior} to another if its branching root is no greater than the other's.
Finally, if every branching rule has branching root at most $c$, then an algorithm using these rules runs in time $O^*(c^d)$, where $d=\mu(I)$ is the measure of the initial instance.
For further details, we refer the reader to the books by Fomin and Kratsch~\cite[Chapter~2]{textbook:FominK10} and Cygan et al.~\cite[Chapter~3]{textbook:CyganFKLMPPS15}.

\section{Our Algorithmic Framework}
\label{sec:framework}

In this section, we present a unified algorithmic framework for computing a $k$-product list for $\ell$-forest deletion sets and $r$-pseudoforest deletion sets.
Concrete instantiations of the framework for these problems are given in \rev{\Cref{sec:lforest,sec:pseudoforest}}.

For clarity, we describe the framework for the problem \delprb, which subsumes \prb{$\ell$-Forest Deletion Set} and \prb{$r$-Pseudoforest Deletion Set}. 
The original input instance is denoted by $(\hat{G}, \hat{k})$, and we use $(G,k)$ or $(G',k')$ for instances arising during the recursive execution of the algorithm.

Our framework consists of two procedures: \ComputingProductList and \Search.
At a high level, \ComputingProductList controls the overall execution of the algorithm, while \Search implements a recursive branching algorithm that constructs $k$-product representations.
In the following subsections, we describe each procedure in detail and explain how they work together to maintain the required invariants and ensure correctness.

\subsection{\texttt{ComputeList} Procedure}

\begin{algorithm}[t]
	\SetAlgoLined
	\SetKwProg{Procedure}{Procedure}{}{ }
	\DontPrintSemicolon
    \KwIn{A multigraph $\hat{G}$ and an integer $\hat{k}$}
	\Procedure{\ComputingProductList{$\hat{G},\hat{k}$}}{
		$\hat{\mathscr{L}} \gets \emptyset$\;
		Compute an $\mathcal{F}$-deletion set $\hat{S}$ of size at most $\hat{k}$ in $\hat{G}$\;
		\revi{\If{no such set exists}{
			\Return $\emptyset$\;
		}}
		\ForEach{$S \subseteq \hat{S}$}{
			$G \gets \hat{G} - S$\;
			Initialize $L \colon V(G) \to 2^{V(\hat{G})}$ with $L(v) \gets \{v\}$ for all $v \in V(G)$\;
            $\hat{\mathscr{L}}_S \gets \Search(G,\hat{k}-|S|,\hat{S}\setminus S,L,\{ \{v\} \mid v \in S \})$\;
            \If{$\hat{\mathscr{L}}_S \neq \emptyset$}{
                $\hat{\mathscr{L}} \gets \hat{\mathscr{L}} \cup \hat{\mathscr{L}}_S$\;
            }
		}
		\Return $\hat{\mathscr{L}}$\;
	}
	\caption{\texttt{ComputeList} procedure}
	\label{alg:computing-product-list}
\end{algorithm}

The framework begins with the procedure \ComputingProductList (see \Cref{alg:computing-product-list}), which computes a $\hat{k}$-product list $\hat{\mathscr{L}}$ of $\mathcal{F}$-deletion sets of the input multigraph $\hat{G}$.
Initially, the list $\hat{\mathscr{L}}$ is empty.
As a preprocessing step, \ComputingProductList computes an $\mathcal{F}$-deletion set $\hat{S}$ of size at most $\hat{k}$ in $\hat{G}$ using an existing fixed-parameter tractable algorithm parameterized by the solution size $\hat{k}$.
If no such set exists, then $\hat{G}$ admits no $\mathcal{F}$-deletion set of size at most $\hat{k}$; in this case, the algorithm returns the empty collection and terminates.
We therefore assume that such a set $\hat{S}$ has been found.
\revi{Importantly, $\hat{S}$ is used only as a reference deletion set and is not assumed to be inclusion-minimal.}

We now introduce the following auxiliary problem.
We are given a multigraph $G$, an integer $k$, a set $F \subseteq V(G)$ with $G - F \in \mathcal{F}$, a labeling function $L \colon V(G) \to 2^{V(\hat{G})}$, and a family $\mathcal{C}$ of pairwise disjoint vertex subsets of $\hat{G}$.
\revi{We maintain the invariant that all sets in $\mathcal{C}\cup\{L(v) \mid v\in V(G)\}$ are nonempty and pairwise disjoint.}
Intuitively, $F$ is a set of undeletable vertices, $L(v)$ represents the set of vertices of $\hat{G}$ that have been identified with $v \in V(G)$ during the execution of \Search, and $\mathcal{C}$ is used to describe components of a $\hat{k}$-product representation of $\hat{G}$ that have already been fixed.
A minimal $\mathcal{F}$-deletion set $\hat{D}$ of $\hat{G}$ is said to be \emph{compatible} with an instance $(G,k,F,L,\mathcal{C})$ of the auxiliary problem if there exists a minimal $\mathcal{F}$-deletion set \revi{$D$} of $G$ satisfying the following three conditions:
\begin{itemize}
\item $|D| \leq k$; and
\item $F \cap D = \emptyset$; and 
\item $\mathcal{C} \cup \{ L(v) \mid v \in D \}$ represents $\hat{D}$. 
\end{itemize}
The goal of the auxiliary problem is to compute a collection $\hat{\mathscr{L}}$ of $\hat{k}$-product representations that represents every minimal $\mathcal{F}$-deletion set of $\hat{G}$ compatible with $(G,k,F,L,\mathcal{C})$.

For each subset $X \subseteq \hat{S}$, the procedure \ComputingProductList invokes the recursive procedure \Search (\Cref{alg:search}), which solves the auxiliary problem for an instance $(G,k,F,L,\mathcal{C})$ initialized as follows.
We set $G \coloneq \hat{G} - X$ and $k \coloneq \hat{k} - |X|$, specify the forbidden vertex set $F$ as $\hat{S} \setminus X$, define a labeling function $L$ by $L(v) \coloneq \{v\}$ for every $v \in V(G)$, and set $\mathcal{C}  \coloneq \bigl\{  \{v\} \mid v \in X \bigr\}$.
Let $\hat{\mathscr{L}}_X$ denote the collection returned by $\Search(G,\hat{k}-|X|,\hat{S}\setminus X,L, \mathcal{C})$.
By definition, every minimal $\mathcal{F}$-deletion set $\hat{D}$ of $\hat{G}$ satisfying $|\hat{D}| \leq \hat{k}$ and $\hat{D} \cap \hat{S} = X$ is represented by some $\hat{k}$-product representation $\mathcal{C}'$ in $\hat{\mathscr{L}}_X$.
Therefore, the union of all collections $\hat{\mathscr{L}}_X$ over all subsets $X \subseteq \hat{S}$ forms a $\hat{k}$-product list of $\mathcal{F}$-deletion sets of $\hat{G}$.

Consequently, in \Cref{sec:lforest,sec:pseudoforest}, it suffices to design reduction and branching rules for \Search that correctly solve the auxiliary problem while preserving the required invariants.

\subsection{\texttt{Search} Procedure}

\begin{algorithm}[t]
	\SetAlgoLined
	\SetKwProg{Procedure}{Procedure}{}{ }
	\DontPrintSemicolon
    \KwIn{A graph $G$, an integer $k$, a subset $F \subseteq V(G)$ such that $G - F \in \mathcal{F}$, a labeling function $L \colon V(G) \to 2^{V(\hat{G})}$, and a family $\mathcal{C}$ of pairwise disjoint vertex subsets of $\hat{G}$}
	\Procedure{\Search{$G,k,F,L,\mathcal{C}$}}{
		\If{a reduction rule applies and yields an instance $(G',k',F',L',\mathcal{C}')$}{
            \Return an output constructed from $\Search(G',k',F',L', \mathcal{C}')$\;
		}
		Apply a branching rule to $(G,k,F,L, \mathcal{C})$ and obtain a list $\mathscr{L}_i$ for each branch $i$\;
		\Return $\bigcup_i \mathscr{L}_i$\;
	}
	\caption{The \texttt{Search} procedure}
	\label{alg:search}
\end{algorithm}

Let $(G,k,F,L,\mathcal{C})$ be an instance of the auxiliary problem.
The procedure \Search is a recursive algorithm that computes partial product representations.
Given such an instance, the procedure first exhaustively executes applicable reduction rules, each of which produces a smaller instance $(G',k',F',L',\mathcal{C}')$ while preserving all minimal $\mathcal{F}$-deletion sets of $G$ of size at most $k$ that are disjoint from $F$.
If no reduction rule is applicable, a branching rule is applied.

Each branching rule generates several subinstances, one for each branch.
\revi{For each branch $i$, a recursive call returns a collection $\mathscr{L}_i$ of product representations, and \Search returns the union $\bigcup_i \mathscr{L}_i$.}
Each branch corresponds to a choice that enforces whether a particular vertex or vertex set is deleted, or instead added to the forbidden set~$F$,
thereby ensuring that all minimal solutions are covered.

The labeling function $L$ encodes the correspondence between vertices of the current graph $G$ and vertices of the original graph $\hat{G}$.
For each vertex $v \in V(G)$, the label $L(v)$ represents the set of vertices of $\hat{G}$ that have been identified with $v$ by reduction rules.
This labeling allows us to apply reduction and branching rules to $G$ while retaining sufficient information to reconstruct product representations for the original graph $\hat{G}$.

\section{Product List of \texorpdfstring{$\ell$}{ell}-Forest Deletion Sets}
\label{sec:lforest}

In this section, we present an algorithm for computing a $k$-product list of $\ell$-forest deletion sets, based on the framework introduced in \Cref{sec:framework}.
To this end, we describe the procedure \Search in detail and introduce several reduction and branching rules.
Our rules are inspired by the FPT algorithm of \cite{lforest:RaiS18} for \prb{$\ell$-Forest Deletion Set};
however, since our objective is different, their correctness must be proved in our setting.

Throughout the execution of \Search, the invariant that $G-F$ is an $\ell$-forest is maintained, since every reduction and branching rule performs only vertex deletions and edge contractions, under which the class of $\ell$-forests is closed.

\subsection{Reductions and Branchings}

In this subsection, we describe the reduction and branching rules used in \Search to compute a list of $\ell$-forest deletion sets.
The first four reduction rules are standard and commonly used in the design of FPT algorithms, for example, for the \prb{Feedback Vertex Set} problem.
In contrast, \Cref{red:deg2_lforest} differs from the standard bypass rule for degree-two vertices.
Instead of removing a single degree-two vertex, we contract a path consisting of two consecutive degree-two vertices, ensuring that the resulting vertex also has degree two.
This modification is crucial for preserving the structure of all minimal $\ell$-forest deletion sets.

Suppose that a reduction rule transforms an instance $(G,k,F,L,\mathcal{C})$ into an instance $(G',k',F',L',\mathcal{C}')$.
The reduction rule is said to be \emph{safe} if every minimal $\mathcal{F}$-deletion set of $\hat{G}$ compatible with $(G,k,F,L,\mathcal{C})$ is also compatible with $(G',k',F',L',\mathcal{C}')$.
Intuitively, safeness ensures that the families $\mathscr{L}$ and $\mathscr{L}'$ obtained by solving $(G,k,F,L,\mathcal{C})$ and $(G',k',F',L',\mathcal{C}')$, respectively, encode the same minimal $\mathcal{F}$-deletion sets of the original graph $\hat{G}$. 

Let $(G,k,F,L,\mathcal{C})$ be an instance of \Search.
We begin with two trivial reduction rules.

\begin{redrule}
    \label{red:no_instance}
    If $k < 0$ or $G[F]$ is not an $\ell$-forest, return \revi{$\emptyset$.}
\end{redrule}

\begin{redrule}
    \label{red:already_solved}
    If $k \geq 0$ and $G$ is an $\ell$-forest, return $\{ \mathcal{C}\}$.
\end{redrule}

We next present three non-trivial reduction rules.

\begin{redrule}
    \label{red:del_degree_one_forest}
    If there exists a vertex $v \in V(G) \setminus F$ with $\deg_G(v) \leq 1$, then return the collection $\Search(G - v, k, F, L|_{G-v},\mathcal{C})$.
\end{redrule}

The safeness of \Cref{red:del_degree_one_forest} follows because a vertex of degree at most~1 lies in no cycle of $G$ and therefore cannot belong to any minimal $\ell$-forest deletion set of $G$.

\begin{redrule}
    \label{red:no_forest}
    If there exists a vertex $v \in V(G) \setminus F$ such that $G[F \cup \{v\}]$ is not an $\ell$-forest, then return the family $\Search(G - v, k - 1, F, L|_{G-v},\mathcal{C}\cup \{L(v)\})$.
\end{redrule}

The safeness follows from the observation that the vertex $v$ must belong to every minimal $\ell$-forest deletion set of $G$ that is disjoint from $F$;
otherwise, $G[F \cup \{v\}]$ would not be an $\ell$-forest, contradicting the definition of a deletion set.

\begin{redrule}
    \label{red:deg2_lforest}
    Let $u_1, u_2$ be adjacent vertices in $G - F$ such that
    $\deg_G(u_1) = \deg_G(u_2) = 2$.
    \revi{Assume that $u_1u_2$ is the unique edge between them.}
    Then return the family $\Search(G / u_1u_2, k, F, L',\mathcal{C})$,
    where $u$ is the vertex obtained by contracting the edge $u_1u_2$, and
    $L'$ is defined by
    \begin{align*}
        L'(v) =
        \begin{cases}
            L(u_1) \cup L(u_2) & \text{if } v = u,\\
            L(v)               & \text{otherwise.}
        \end{cases}
    \end{align*}
\end{redrule}

\begin{lemma}
\label{lem:safeness_deg2}
\Cref{red:deg2_lforest} is safe.
\end{lemma}

\begin{proof}
Let $\hat{D}$ be a minimal $\ell$-forest deletion set of $\hat{G}$ that is compatible with $(G,k,F,L,\mathcal{C})$.
By definition, there exists a minimal $\mathcal{F}$-deletion set $D$ of $G$ with size at most $k$ such that $F \cap D = \emptyset$ and 
$\mathcal{C} \cup \{ L(v) \mid v \in D \}$ represents $\hat{D}$.
We claim that $\hat{D}$ is also compatible with $(G / u_1u_2, k, F, L',\mathcal{C})$.

If $u_1, u_2 \notin D$, then $D$ is a minimal $\ell$-forest deletion set of $G / u_1u_2$, since contracting the non-loop edge $u_1u_2$ preserves the excess of $G - Z$ for every $Z \subseteq D$.
Thus, in this case, $\hat{D}$ is compatible with $(G / u_1u_2, k, F, L',\mathcal{C})$.

Next, suppose that exactly one of $u_1$ and $u_2$ belongs to $D$.
Note that $D$ does not contain both $u_1$ and $u_2$, since $u_1 \in D$ and $u_2 \in D$ imply that $D$ is not minimal.
Without loss of generality, assume that $u_1 \in D$.
Then let $D' = (D \setminus \{u_1\}) \cup \{ u\}$.
Since $\deg_G(u_1) = \deg_G(u_2) = 2$, for every $Z \subseteq D \setminus \{u_1\}$, the graph $(G / u_1u_2) - (Z \cup \{u\})$ is obtained from $G - (Z \cup \{u_1\})$ by deleting the vertex $u_2$ of degree at most one, which preserves the excess.
Together with the previous case applied to subsets of $D \setminus \{u_1\}$, it follows that $D'$ is a minimal $\ell$-forest deletion set of $G / u_1u_2$.
Moreover, since $u \notin F$, we have $F \cap D' = \emptyset$.
By definition, we have $L(v) = L'(v)$ for every $v \in D'\setminus \{u\}$ and $L(u_1) \subseteq L'(u)$.
Consequently, $\mathcal{C} \cup \{ L'(v) \mid v \in D' \}$ also represents $\hat{D}$, which shows that $\hat{D}$ is compatible with $(G / u_1u_2, k, F, L',\mathcal{C})$.
\end{proof}

We consider the following two branching rules.

\begin{branching}
    \label{branching:lforest_standard}
    If there exists a connected set $U \subseteq V(G) \setminus F$ with
    $\deg_{F}(U) \geq 2$ and $|U| \leq 3$,
    then we choose such a set $U$ of minimum cardinality and branch along a path in $G[U]$, starting from a vertex adjacent to $F$.
    Formally, let $u_1, \ldots, u_{|U|}$ be the vertices of $U$ ordered along such a path in $G[U]$, where $\deg_F(u_1) \geq 1$, and let $U_i \coloneqq \{u_1, \ldots, u_i\}$ for each $i \in\{0, 1, \ldots, |U|\}$, setting $U_0 \coloneqq \emptyset$.
    The branching generates the following instances:
    \begin{align*}
        &I_i = (G - u_i, k - 1, F \cup U_{i-1}, L|_{G - u_i},\mathcal{C}\cup \{L(u_i)\}) ~\text{for each $i \in [|U|]$, and} &\\
        &I_{|U|+1} = (G, k,F \cup U, L,\mathcal{C}),&
    \end{align*}
    where, for a subgraph $G'$ of $G$, $L|_{G'}$ denotes the restriction of $L$ to $V(G')$.
\end{branching}

\begin{branching}
    \label{branching:lforest_cycle}
    If the graph $G - F$ contains a cycle, choose an arbitrary vertex $v$ on such a cycle
    and branch on 
    \begin{align*}
        I_1 = (G - v, k - 1, F, L|_{G - v},\mathcal{C}\cup \{L(v)\}) \quad \text{and} \quad
        I_2 = (G, k,F \cup \{v\}, L,\mathcal{C}).
    \end{align*}
\end{branching}

It is easy to see the following observation; otherwise, we can apply some reduction rules and \Cref{branching:lforest_standard} to the instance.

\begin{observation}
\label{obs:red_after_deg}
Let $(G,k,F,L,\mathcal{C})$ be an input of \Search.
If none of \Cref{red:no_instance,red:already_solved,red:no_forest,red:deg2_lforest,red:del_degree_one_forest,branching:lforest_standard} applies, then the following properties hold:
\begin{enumerate}
    \item For every vertex $v\in V(G)\setminus F$, we have $\deg_G(v)\ge 2$.
    \item If a vertex $v\in V(G)\setminus F$ satisfies $\deg_{G-F}(v)=1$, then its unique neighbor
    $u$ satisfies $\deg_G(u)\ge 3$.
\end{enumerate}    
\end{observation}

\begin{proof}
    \revi{The first claim follows from the fact that we cannot apply \Cref{red:del_degree_one_forest}.
    For the second claim, suppose that $\deg_{G-F}(v)=1$ and let $u$ be its neighbor in $G-F$.
    Since \Cref{branching:lforest_standard} does not apply to the singleton $\{v\}$, we have $\deg_F(v) \leq 1$.
    Thus, the first claim gives $\deg_G(v)=2$.
    If $\deg_G(u)=2$, then \Cref{red:deg2_lforest} applies.
    Thus $\deg_G(u)\ge3$.}
\end{proof}

We observe that the above two branching rules are exhaustive.

\begin{lemma}
\label{lem:l_forest_branching}
If \Cref{red:no_instance,red:already_solved,red:no_forest,red:deg2_lforest,red:del_degree_one_forest,branching:lforest_standard} are not applicable, then $G - F$ contains a cycle.
Equivalently, \Cref{branching:lforest_cycle} is applicable.
\end{lemma}

\begin{proof}
We prove the claim by contradiction.
Suppose that $G - F$ is acyclic,
that is, every connected component of $G - F$ is a tree.

Let $C$ be an arbitrary connected component of $G - F$, and let $T = G[C]$.
Since no reduction rules or \Cref{branching:lforest_standard} apply,
every connected component of $G - F$ that has a neighbor in $F$ contains at least four vertices. 
Thus, we have $|V(T)| \geq 4$.

Let $P$ be a path of $T$ attaining the diameter of $T$, and let $v$ be an endpoint of $P$.
By the choice of $P$, the vertex $v$ is a leaf of $T$, and hence $\deg_{G-F}(v)=1$.
By Claim~(1) of \Cref{obs:red_after_deg}, we have $\deg_G(v)\ge 2$, which implies that $v$ has a neighbor in $F$.
Since \Cref{branching:lforest_standard} is not applicable, $v$ cannot have more than one neighbor in $F$.
Therefore, $v$ is adjacent to exactly one vertex in $F$.

Let $v_1$ be the unique neighbor of $v$ in $T$.
By Claim~(2) of \Cref{obs:red_after_deg}, the degree of $v_1$ in $T$ is at least $3$.
Note that $v_1$ has no neighbor in $F$; otherwise, \Cref{branching:lforest_standard} is applicable.
Thus, $v_1$ has a neighbor $v_2$ in $T$ that does not lie on the path $P$.
Since $T$ is a tree and $P$ is a diameter-attaining path, the vertex $v_2$ must be a leaf of $T$.
Moreover, Claim~(1) of \Cref{obs:red_after_deg} implies that 
the vertex $v_2$ is adjacent to a vertex in $F$.

Consequently, the connected set $U \coloneq \{v, v_1, v_2\} \subseteq V(G) \setminus F$
satisfies $|U| \leq 3$ and $\deg_{F}(U) \geq 2$.
Thus, \Cref{branching:lforest_standard} is applicable, contradicting the assumption that this branching rule cannot be applied.
\end{proof}

\subsection{Running Time Analysis}
To analyze the running time of \Search, we define the following measure for an input instance $I = (G, k, F, L,\mathcal{C})$ of \Search as
\begin{align*}
    \mu(I) = k + \alpha \cdot \contcomp{G[F]} + \beta \cdot \excess{G -F} + \bigl(\ell - \excess{G[F]}\bigr),
\end{align*}
where $\alpha$ and $\beta$ are positive constants to be specified later.
This measure has been used previously to design FPT algorithms for \prb{$\ell$-Forest Deletion Set}~\cite{lforest:LinFWCFL18,lforest:RaiS18}.
Observe that none of the reduction rules increases the value of the measure.

\medskip
\noindent
\textbf{\Cref{branching:lforest_standard}.}
We only consider the case $|U| = 3$, as the other cases can be handled analogously and the case $|U| = 3$ is the worst case.
Let $U = \{u_1,u_2,u_3\} \subseteq V(G)\setminus F$ be a connected set satisfying the conditions of
\Cref{branching:lforest_standard}.
This branching rule produces the following instances:
\begin{align*}
    &I_i = (G - u_i, k - 1, F \cup U_{i-1}, L|_{G - u_i},\mathcal{C}\cup \{L(u_i)\}) ~\text{for each $i \in [|U|]$, and} &\\
    &I_{|U|+1} = (G, k,F \cup U, L,\mathcal{C}).&
\end{align*}

In the instances $I_1$, $I_2$, and $I_3$, the parameter $k$ decreases by one, and none of the remaining terms in the measure increases.
Thus, the measures for these instances decrease by at least one.

In the instance $I_4$, the parameter $k$ remains unchanged.
Since $\deg_{F}(U) \geq 2$ and $G[U]$ is connected, $\contcomp{G[F]}$ does not increase.
If $\contcomp{G[F]}$ decreases, then the measure decreases by at least $\alpha$.
Otherwise, $\contcomp{G[F]}$ remains unchanged, and since $\excess{G[F]}$ increases by at least one, the term $\ell - \excess{G[F]}$ decreases by at least one.
Thus, the measure decreases by at least $\min\{1, \alpha\}$.

Therefore, the branching vector for \Cref{branching:lforest_standard} is $(1, 1, 1, \min\{1, \alpha\})$.

\medskip
\noindent
\textbf{\Cref{branching:lforest_cycle}.}
Let $v$ be a vertex that lies on a cycle in $G - F$.
This branching rule produces the following two instances:
\begin{align*}
    I_1 = (G - v, k - 1, F, L|_{G - v},\mathcal{C}\cup \{L(v)\})
    \quad \text{and} \quad
    I_2 = (G, k,F \cup \{v\}, L,\mathcal{C}).
\end{align*}

In the instance $I_1$, the parameter $k$ decreases by one, while the other terms of the measure do not increase.
In the instance $I_2$, the parameter $k$ remains unchanged and $\contcomp{G[F]}$ increases by at most one.
By \Cref{fds:subgraph}, the quantity $\excess{G - F}$ decreases by at least one, which implies a decrease of at least $\beta$ in the measure.
Moreover, since \Cref{branching:lforest_standard} is not applicable, $v$ has at most one neighbor in $F$, and hence $\excess{G[F]}$ does not decrease.
Thus, the term $\ell - \excess{G[F]}$ does not increase.
Overall, the measure decreases by at least $\beta - \alpha$.

Therefore, the branching vector for \Cref{branching:lforest_cycle} is $(1, \beta - \alpha)$.
By choosing constants $\alpha$ and $\beta$ such that $\beta > \alpha$ (e.g., $\alpha = 1$ and $\beta = 2$), the measure strictly decreases in every branching step.

\medskip
\noindent
\textbf{Running Time and Correctness.}
We analyze the running time of \ComputingProductList.
Let $I = (\hat{G} - S, \hat{k} - |S|, \hat{S} \setminus S, L,\{ \{v\} \mid v \in S\})$ be an instance of \Search invoked during the execution of
\ComputingProductList.
For this instance, $\contcomp{G[F]}\leq |\hat S\setminus S|\leq \hat k-|S|$, $\excess{G-F}\leq\ell$, and $\excess{G[F]}\ge0$.  
Therefore,
\[
\mu(I) \leq (1+\alpha)(\hat{k} - |S|) + (\beta+1)\ell.
\]
Setting $\alpha = 1$ and $\beta = 2$, the worst branching vector is $(1,1,1,1)$.\footnote{We first use $\alpha=1$ and $\beta=2$ to obtain a simple bound; optimized values are considered below.}
Thus, the running time of \Search on instance $I$ is bounded by $4^{\mu(I)} \cdot n^{O(1)}$.
Substituting the bound on $\mu(I)$, we obtain $4^{2(\hat{k} - |S|) + 3\ell} \cdot n^{O(1)}$.

The algorithm considers every subset $S \subseteq \hat{S}$.
Therefore, the total running time of \ComputingProductList is
\begin{align*}
    \sum_{S \subseteq \hat{S}}
    4^{2(\hat{k} - |S|) + 3\ell} \cdot n^{O(1)}
    \leq 64^{\ell} \cdot \sum_{i=0}^{\hat{k}} \binom{\hat{k}}{i} 4^{2(\hat{k}-i)} \cdot n^{O(1)} 
    = 64^{\ell} \cdot 17^{\hat{k}} \cdot n^{O(1)}.
\end{align*}

The correctness of the algorithm follows from the safeness of all reduction rules, from the fact that some branching rule is always applicable, as established in \Cref{lem:l_forest_branching}, and from the fact that the branching rules exhaustively cover all feasible cases: a minimal $\mathcal{F}$-deletion set $D$ either avoids all vertices branched on, or $D\setminus\{u_i\}$ is a minimal $\mathcal{F}$-deletion set of $G-u_i$ for the first such vertex $u_i$ in $D$.
This completes the proof of the following result.

\begin{theorem}
    \label{thm:lforest_comp_rep}
    Let $(G, k)$ be an instance of \prb{$\ell$-Forest Deletion Set}.
    There is an algorithm computing a $k$-product list for $\ell$-forest deletion sets of $G$, in FPT time with respect to $\ell + k$.    
\end{theorem}

By setting $\ell = 0$, we obtain the following corollary.

\begin{corollary}
    \label{cor:fvs_comp_rep}
    Let $(G, k)$ be an instance of \prb{Feedback Vertex Set}.
    There is an algorithm computing a $k$-product list for feedback vertex sets of $G$ in time $17^k \cdot n^{O(1)}$.        
\end{corollary}

To obtain improved running-time bounds, we use the measure-and-conquer technique~\cite{FominGK09}.
Setting $\alpha = \log_6{2}$ and $\beta = \log_6{\frac{12}{5}}$ yields a running time of $O^*(13^k \cdot 14.4^\ell)$.
Alternatively, setting $\alpha = 0.317014053$ and $\beta = 0.403115906$ yields a running time of $O^*(14.27^{k+\ell})$.
The source code used to compute these bounds is available at \url{https://github.com/srin728/ProductList_analysis}.

\section{Product Lists for \texorpdfstring{$r$}{r}-Pseudoforest Deletion Sets} \label{sec:pseudoforest}

In this section, we present an algorithm for computing a $k$-product list of $r$-pseudoforest deletion sets, based on the framework introduced in \Cref{sec:framework}.
Most reduction rules and the first branching rule are identical to those presented in \Cref{sec:lforest}; in addition, we introduce one new reduction rule and four branching rules specific to $r$-pseudoforests.
The safeness of the reduction rules follows from the same arguments as in \Cref{sec:lforest}.
Some of our branching rules are inspired by the work of Sheng and Gutin~\cite{rpseudoforest:ShengG23} and Tsur~\cite{rpseudoforest:Tsur25}.
\revi{Unlike the case of \prb{$\ell$-Forest Deletion Set}, however, the existing branching rules cannot be extended to \prb{$r$-Pseudoforest Deletion Set} in a straightforward manner.
To overcome this difficulty, we introduce an auxiliary procedure, which serves as a key ingredient in the design of our branching algorithm.}

In this section, we assume that $r \geq 1$, since the case $r = 0$ (feedback vertex sets) is handled in \Cref{cor:fvs_comp_rep}.

\subsection{Reduction and Branching Rules}
Let $(G,k,F,L,\mathcal{C})$ be an instance of \Search.
We first introduce the following reduction rules, where \Cref{red:no_instance_r,red:already_solved_r,red:del_degree_one_rforest} play the same role as \Cref{red:no_instance,red:already_solved,red:del_degree_one_forest}, respectively.

\begin{redrule}
    \label{red:no_instance_r}
    If $k < 0$ or $G[F]$ is not an $r$-pseudoforest, return $\emptyset$.
\end{redrule}

\begin{redrule}
    \label{red:already_solved_r}
    If $k \geq 0$ and $G$ is an $r$-pseudoforest, return $\{ \mathcal{C}\}$.
\end{redrule}

\begin{redrule}
    \label{red:forest_comp}
    If a connected component $C$ of $G - F$ satisfies $\deg_{F}(C) = 0$, then return the collection $\Search(G - C, k, F, L|_{G-C},\mathcal{C})$.
\end{redrule}

\begin{redrule}
    \label{red:del_degree_one_rforest}
    If there exists a vertex $v \in V(G) \rev{\setminus F}$ with $\deg_G(v) \leq 1$, then return the collection $\Search(G - v, k, F, L|_{G-v},\mathcal{C})$.
\end{redrule}

The safeness of \Cref{red:no_instance_r,red:already_solved_r,red:del_degree_one_rforest} is immediate.
The safeness of \Cref{red:forest_comp} follows from the fact that $G[C]$ forms an $r$-forest in $G$; hence, no vertex of $C$ is contained in any minimal $r$-pseudoforest deletion set.

The following reduction rule is identical to \Cref{sec:lforest} and the safeness follows by the same argument.

\begin{redrule}
    \label{red:deg2_rforest}
    Let $u_1, u_2$ be adjacent vertices in $G - F$ such that
    $\deg_G(u_1) = \deg_G(u_2) = 2$.
    Assume that $u_1u_2$ is the unique edge between them.
    Then return the family $\Search(G / u_1u_2, k, F, L',\mathcal{C})$,
    where $u$ is the vertex obtained by contracting the edge $u_1u_2$, and
    $L'$ is defined by
    \begin{align*}
        L'(v) =
        \begin{cases}
            L(u_1) \cup L(u_2) & \text{if } v = u,\\
            L(v)               & \text{otherwise.}
        \end{cases}
    \end{align*}
\end{redrule}

We also reuse the branching rule used for $\ell$-forest deletion (see \Cref{sec:lforest}).

\begin{branching}
    \label{branching:rpseudo_forest_standard}
    If there exists a connected set $U \subseteq V(G) \setminus F$ such that
    $\deg_{F}(U) \geq 2$ and $|U| \leq 3$,
    choose such a set $U$ of minimum cardinality and branch along a path in $G[U]$ starting at a vertex adjacent to $F$.
    Formally, let $u_1, \ldots, u_{|U|}$ be the vertices of $U$ ordered along such a path, with $\deg_F(u_1)\ge1$. Define $U_0\coloneq\emptyset$ and $U_i\coloneq\{u_1,\ldots,u_i\}$ for $i\in[|U|]$.
    The branching generates instances, defined as follows:
    \begin{align*}
        &I_i = (G - u_i, k - 1, F \cup U_{i-1}, L|_{G - u_i},\mathcal{C}\cup \{L(u_i)\}) ~\text{for each $i \in [|U|]$, and} &\\
        &I_{|U|+1} = (G, k,F \cup U, L,\mathcal{C}),&
    \end{align*}
    where, for a subgraph $G'$ of $G$, $L|_{G'}$ denotes the restriction of $L$ to $V(G')$.
\end{branching}

The following lemma is central to our branching rule.

\begin{lemma}
\label{lem:pseudo_forest_key}
    Let $T$ be a tree and let $s$ be a nonnegative integer.
    Let $L, Q \subseteq V(T)$ with $|L| \leq 2s$ and $|Q| \geq 2$.
    Suppose that every leaf of $T$ lies in $L \cup Q$, and that no two distinct vertices in $V(T) \setminus (L \cup Q)$ of degree~$2$ are adjacent.
    Then there exist distinct vertices $u, v \in Q$ such that $\dist{T}{u}{v} \leq 4s+4$.
\end{lemma}

\begin{proof}
Let $V_{\geq 3} \coloneq \{v \in V(T) \mid \deg_T(v) \geq 3\}$ and $X \coloneq L \cup Q \cup V_{\geq 3}$.

For a graph $G$ and a vertex $v \in V(G)$ of degree~$2$, we \emph{suppress} $v$ by deleting $v$ and adding an edge between its neighbors.
Let $T'$ be the tree obtained from $T$ by suppressing every vertex of degree~$2$ in $V(T) \setminus (L \cup Q)$.
By construction, $V(T') = X$.
Moreover, since suppressing every vertex of degree~$2$ does not create new leaves, every leaf of $T'$ also belongs to $L \cup Q$.
Since no two vertices in $V(T) \setminus (L \cup Q)$ of degree~$2$ are adjacent, each edge of $T'$ corresponds to a path in $T$ of length at most~$2$. 
Thus, for any $x, y \in X$,
\begin{equation}
    \label{eq:tree_distance}
    \dist{T}{x}{y} \leq 2 \cdot \dist{T'}{x}{y}.
\end{equation}

Let $T^*$ be the minimal subtree of $T'$ that contains all vertices in $Q$. 
By minimality, every leaf of $T^*$ belongs to $Q$, and since $|Q| \geq 2$, we have $|V(T^*)| \geq 2$.

We first show that there exist two vertices $q_1, q_2 \in Q$ such that the $q_1$--$q_2$ path $P$ in $T^*$ has no internal vertex in $Q$ and contains at most one vertex of degree at least~$3$.
If $T^*$ is a path, the claim is immediate.
Otherwise, consider $T^*$ rooted at an arbitrary vertex, and let $w$ be a vertex of degree at least~$3$ that is farthest from the root. 
Then two distinct descendants of $w$ define paths to leaves that do not contain any other vertex of degree at least~$3$. 
Let $P$ be a path in the subtree rooted at $w$ whose endpoints lie in $Q$ and whose internal vertices contain no vertices of $Q$.
Let $q_1$ and $q_2$ be endpoints of $P$.

Let $I$ be the set of internal vertices of $P$, and let $I' \subseteq I$ be the set of vertices of degree~$2$ in $T^*$.
By the choice of $q_1$ and $q_2$, we have $I \cap Q = \emptyset$.
Since $P$ contains at most one vertex of degree at least~$3$ in $T^*$, it follows that $|I| \leq |I'| + 1$.

We bound $|I'|$ by constructing an injection from $I'$ to $L$.
Fix $v \in I'$.
Since $v \in X \setminus Q$, either $v \in L$ or $\deg_T(v) \geq 3$.

If $v \in L$, map $v$ to itself.
Otherwise, $\deg_T(v) \geq 3$, and hence $\deg_{T'}(v) \geq 3$.
Since $v \in I'$ has degree~$2$ in $T^*$, it has a neighbor in $T'$ that does not belong to $T^*$. 
By following such an edge and proceeding away from $T^*$, we reach a leaf $\ell$ of $T'$.
As $\ell \notin Q$, we have $\ell \in L$, and we map $v$ to $\ell$.

These mappings are injective, since distinct vertices in $I'$ correspond to distinct vertices in $L$.
Therefore, $|I'| \leq |L| \leq 2s$.

Consequently, $|I| \leq |I'| + 1 \leq 2s+1$, and thus $\dist{T'}{q_1}{q_2} \leq 2s+2$.
By \Cref{eq:tree_distance},
\[
\dist{T}{q_1}{q_2} \leq 2\,\dist{T'}{q_1}{q_2} \leq 4s+4,
\]
which completes the proof.
\end{proof}

\begin{algorithm}[t]
	\SetAlgoLined
	\SetKwProg{Procedure}{Procedure}{}{ }
	\DontPrintSemicolon
    \KwIn{A graph $G$, a set $F \subseteq V(G)$ such that $G - F$ is an $r$-pseudoforest, and a connected component $C$ of $G - F$}
	\Procedure{\textsc{ContSet}}{
        choose a set $E' \subseteq E(G[C])$ with $|E'| \leq r$ such that $G[C] - E'$ is a tree\;
        $T \gets G[C] - E'$\;
		$Q_C \gets \{ u \in C \mid \deg_F(u) = 1 \}$\;
        $L \gets$ the set of endpoints of edges in $E'$\;
        
        initialize $T' \gets T$, $Q' \gets Q_C$, and $L' \gets L$\;
        initialize a partition $\mathcal{B} = \{V_w\}_{w \in V(T')}$ with $V_w = \{w\}$ for every $w \in V(T')$\;
        
        define $\Qrank_{T'}(w) \coloneqq |V_w \cap Q_C|$ for each $w \in V(T')$\;
        define $\lab_{T'}(w) \coloneqq i$ for each $w \in Q'$ adjacent to a vertex in $R_i$\;
        
        \While{$|Q'| \geq 2$ and $\Qrank_{T'}(w) \leq r+1$ for every $w \in V(T')$}{
            choose $u,v \in Q'$ that minimize $\dist{T'}{u}{v}$ \tcp*{$\dist{T'}{u}{v} \leq 4r + 4$}
            $P \gets$ the $u$--$v$ path in $T'$ \tcp*{$V(P) \cap Q' = \{u,v\}$}
                        
            construct $T''$ from $T'$ by contracting all edges of $P$ into a new vertex $w_{uv}$\;
            
            define $V_{w_{uv}} \coloneqq \bigcup_{x \in V(P)} V_x$\;
            
            \If{$\lab_{T'}(u) \neq \lab_{T'}(v)$}{
                \Return $V_{w_{uv}}$\;
                \label{line:distinct_comp}
            }
            update $\mathcal{B} \gets \{V_x \mid x \in V(T') \setminus V(P)\} \cup \{V_{w_{uv}}\}$\;
            
            $Q' \gets (Q' \setminus V(P)) \cup \{w_{uv}\}$\;
            $L' \gets L' \setminus V(P)$\;

            define $\Qrank_{T''}(w_{uv}) \coloneqq |V_{w_{uv}} \cap Q_C| \ (= \Qrank_{T'}(u) + \Qrank_{T'}(v))$ and set $\Qrank_{T''}(x) \coloneqq \Qrank_{T'}(x)$ for each $x \in V(T') \setminus V(P)$\; 
            define $\lab_{T''}(w_{uv}) \coloneqq \lab_{T'}(u)$ and set $\lab_{T''}(x) \coloneqq \lab_{T'}(x)$ for each $x \in Q' \setminus \{w_{uv}\}$\;
        
            $T' \gets T''$\;
        }
        \Return a set $V_w \in \mathcal{B}$ maximizing $\Qrank_{T'}(w)$\;
	}
	\caption{$\textsc{ContSet}$ procedure}
	\label{alg:cont_set}
\end{algorithm}

Let $C$ be a connected component of $G - F$.
After exhaustively applying \Cref{branching:rpseudo_forest_standard}, every vertex $u \in C$ satisfies $\deg_F(u) \leq 1$.
Let $Q_C \coloneq \{ u \in C \mid \deg_F(u) = 1 \}$.

We now consider the case that $|Q_C| \geq 2$.
Our strategy is to apply \Cref{alg:cont_set} together with \Cref{lem:pseudo_forest_key} to obtain a connected set $C' \subseteq C$ such that both $|C'|$ and $\deg_F(C')$ are bounded by a function of $r$.
We then apply one of the branching rules in \Cref{branching:shortest_path,branching:naive,branching:connected}, depending on the value of $\deg_F(C')$ and the structure of $N_F(C')$.
If $|Q_C|=1$, we instead apply \Cref{branching:one_cc}; we apply this rule only when no other branching rule is applicable.

We briefly describe \Cref{alg:cont_set}.
Since $G-F$ is an $r$-pseudoforest, the connected graph $G[C]$ is an $r$-forest.
Thus, there exists a set $E' \subseteq E(G[C])$ with $|E'| \leq r$ such that $T \coloneqq G[C] - E'$ is a tree.
Let $L$ be the set of endpoints of edges in $E'$.

The procedure maintains a tree $T'$, a partition $\mathcal{B} = (V_w)_{w \in V(T')}$ of $C$, and subsets $Q', L' \subseteq V(T')$.
The tree $T'$ is obtained from $T$ by repeatedly contracting edges, and $\mathcal{B}$ records the partition of $C$ induced by these contractions.
Equivalently, there exists a minor model $\phi$ of $T'$ in $T$ such that $\mathcal{B} = \{ \phi(w) \mid w \in V(T') \}$.
The sets $Q'$ and $L'$ correspond to vertices in $T'$ representing $Q_C$ and $L$, respectively.
For each $w\in V(T')$, we define $\Qrank_{T'}(w)\coloneqq |V_w\cap Q_C|$.
Let $R_1, R_2, \ldots, R_{\cc(G[F])}$ be the connected components of $G[F]$ under a fixed ordering. 
Initially, we define a labeling function $\lab_{T'}\colon Q'\rightarrow[\cc(G[F])]$ by setting $\lab_{T'}(v)=i$ if $v\in Q'$ is adjacent to a vertex of $R_i$.
Since \Cref{branching:rpseudo_forest_standard} is not applicable, every vertex of $Q_C$ has exactly one neighbor in $F$, and therefore $\lab_{T'}$ is well-defined.

In each iteration, the algorithm chooses vertices $u,v\in Q'$ minimizing $\dist{T'}{u}{v}$.
If $|Q'| \geq 2$, such a pair satisfies $\dist{T'}{u}{v} \leq 4r+4$ by \Cref{lem:pseudo_forest_key}.
Indeed, the assumptions of \Cref{lem:pseudo_forest_key} hold with $s=r$, $L=L'$, and $Q=Q'$.
Initially, we have $|L|\leq 2r$, a leaf of $T$ outside $L$ has a neighbor in $F$ since \Cref{red:del_degree_one_rforest} does not apply, and two adjacent vertices of degree~$2$ in $T$ outside $L\cup Q_C$ would have degree~$2$ in $G$, so \Cref{red:deg2_rforest} would apply.
These properties are preserved, since $w_{uv}$ belongs to $Q'$ and every vertex outside $P$ has at most one neighbor on $P$ in the tree $T'$, which implies that its degree does not change.

Let $P$ be the $u$--$v$ path in $T'$.
The algorithm contracts all edges of $P$ into a single vertex and updates $T'$, $Q'$, $L'$, and $\mathcal{B}$ accordingly.
The procedure continues while $|Q'| \geq 2$ and $\Qrank_{T'}(w)\leq r+1$ for every vertex $w\in V(T')$.
Finally, it outputs a set $V_w \in \mathcal{B}$ that maximizes $\Qrank_{T'}(w)$.

We first observe that \Cref{alg:cont_set} runs in polynomial time.
Moreover, the output set $V_w$ is connected.
We next bound the size of $V_w$.

\begin{lemma}
\label{lem:output_size}
In each iteration, for every $w \in V(T'')$ with $i=\Qrank_{T''}(w)\ge 1$, it holds that
\[
|V_w| \leq (4r + 3)(i - 1) + i.
\]
\end{lemma}

\begin{proof}
    The claim is trivial for $i = 1$, so assume that $i \geq 2$.
    
    Let $w_{uv}$ be the vertex obtained by contracting the path $P$ between two vertices $u$ and $v$. 
    Suppose that $\Qrank_{T''}(w_{uv}) = i$ after the contraction.
    Let $\Qrank_{T'}(u) = j$ and $\Qrank_{T'}(v) = i - j$ for some $1 \leq j < i$. 
    By construction, the path $P$ has length at most $4r+4$.

    Since $u$ and $v$ are chosen as the closest pair of vertices in $Q'$, no internal vertex of $P$ belongs to $Q'$. 
    Moreover, by the construction of the algorithm, every vertex $x \in V(T') \setminus Q'$ satisfies $|V_x| = 1$.
    Therefore,
    \[
    \left|\bigcup_{x \in V(P) \setminus \{u,v\}} V_x \right|
    = \sum_{x \in V(P) \setminus \{u,v\}} |V_x|
    = |V(P)| - 2 \leq (4r + 5) - 2 = 4r + 3.
    \]

    By the induction hypothesis, $|V_u| \leq (4r+3)(j-1)+j$ and $|V_v| \leq (4r+3)(i-j-1)+(i-j)$.
    Since $V_{w_{uv}}$ is obtained by taking the union of $V_u$, $V_v$, and the sets corresponding to the internal vertices of $P$, we obtain
    \begin{align*}
    |V_{w_{uv}}|
    &\leq |V_u| + |V_v| + \left|\bigcup_{x \in V(P) \setminus \{u,v\}} V_x \right| \\
    &\leq (4r+3)(j-1)+j + (4r+3)(i-j-1)+(i-j) + (4r+3) \\
    &= (4r+3)(i-1) + i,
    \end{align*}
    which completes the proof.
\end{proof}

Since $\Qrank_{T'}(w) \leq r+1$ holds throughout the execution of the while-loop, and each contraction merges exactly two vertices of $Q'$, every vertex satisfies $\Qrank_{T'}(w) \leq 2r+2$ upon termination.

\begin{observation}
\label{obs:output_rank}
    Let $V_w$ be the output set of \Cref{alg:cont_set}.
    Then $\Qrank_{T'}(w) \leq 2r+2$, and hence $|V_w| \leq (4r+3)(2r+1) + (2r+2) = 8r^2+12r+5$ by \Cref{lem:output_size}.
\end{observation}

Let $C'$ be the connected set returned by \Cref{alg:cont_set} when applied to a connected component $C$. 
In the following, we apply \Cref{branching:shortest_path,branching:naive,branching:connected} according to the output set $C'$. 
We first consider the case where Line~15 of \Cref{alg:cont_set} is executed.

\begin{branching}
    \label{branching:shortest_path}
    Let $C$ be a connected component of $G - F$ such that $C'$ is adjacent to at least two distinct connected components of $G[F]$. 
    Let $P = (v_1, v_2, \ldots, v_t)$ be a shortest path in $G[C']$ such that $v_1$ and $v_t$ have neighbors in different connected components of $G[F]$.
    The branching generates instances, defined as follows:
    \begin{align*}
        &I_i =
        (G - v_i, k - 1, F \cup \{v_1, \ldots, v_{i-1}\}, L|_{G - v_i},\mathcal{C}\cup \{L(v_i)\})
        \quad \text{for each } i \in [t], \text{ and}\\
        &I_{t+1} = (G, k, F \cup \{v_1, \ldots, v_t\}, L,\mathcal{C}).
    \end{align*}
\end{branching}

We next consider the case where Line~15 of \Cref{alg:cont_set} is not executed.
Observe that if $2 \leq |Q_C| \leq r + 1$ (that is, $\deg_F(C) \in [2, r+1]$), then $C'$ contains all vertices of $Q_C$.
We therefore apply the following two branching rules.

\begin{branching}
    \label{branching:naive}
    Let $C$ be a connected component of $G - F$ with $\deg_F(C) \in [2, r+1]$ such that $C'$ is adjacent to exactly one connected component of $G[F]$.
    Let $(v_1, v_2, \ldots, v_{|C'|})$ be a depth-first search ordering of $C'$, where $v_1 \in Q_C$.
    The branching generates instances, defined as follows:
    \begin{align*}
        &I_i =
        (G - v_i, k - 1, F \cup \{v_1, \ldots, v_{i-1}\}, L|_{G - v_i},\mathcal{C}\cup \{L(v_i)\})
        \quad \text{for each } i \in [|C'|], \text{ and}\\
        &I_{|C'|+1} = (G, k, F\cup C', L,\mathcal{C}).
    \end{align*}
\end{branching}

We next handle the case where $\deg_F(C) \geq r+2$.

\begin{branching}
    \label{branching:connected}
    Let $C$ be a connected component of $G - F$ with $\deg_F(C) \geq r + 2$ such that $C'$ is adjacent to exactly one connected component of $G[F]$.
    Let $(v_1, v_2, \ldots, v_{|C'|})$ be a depth-first search ordering of $C'$, where $v_1 \in Q_C$.
    The branching generates instances, defined as follows:
    \begin{align*}
    I_i =
    (G - v_i, k - 1, F \cup \{v_1, \ldots, v_{i-1}\},L|_{G - v_i},\mathcal{C}\cup \{L(v_i)\})
    \quad \text{for each } i \in [|C'|].        
    \end{align*}
\end{branching}

Recall from \Cref{obs:output_rank} that the number of instances generated by \Cref{branching:connected} is bounded by a function of $r$.
We next prove the safeness of \Cref{branching:connected}.

\begin{lemma} \label{lem:comp_forb}
    In \Cref{branching:connected}, every $r$-pseudoforest deletion set of $G$ that is disjoint from $F$ intersects $C'$.
\end{lemma}

\begin{proof}
    Since \Cref{branching:shortest_path} does not apply, $C'$ is adjacent to a unique connected component $R$ of $G[F]$.
    Let $H \coloneq G[C' \cup R]$.
    Since $G[C']$ is connected, we have $|E(G[C'])| \geq |C'| - 1$.
    Similarly, $|E(G[R])| \geq |R| - 1$.
    By construction, there are at least $r+2$ edges between $C'$ and $R$.
    Thus,
    \begin{align*}
        |E(H)|
        &\geq |E(G[C'])| + |E(G[R])| + (r+2) \\
        &\geq (|C'| - 1) + (|R| - 1) + (r+2) \\
        &= |C'| + |R| + r \\
        &= |V(H)| + r.
    \end{align*}
    Hence, $H$ is not an $r$-pseudoforest.
    This implies that every $r$-pseudoforest deletion set of $G$ \rev{that is disjoint from $F$} intersects $C'$.
\end{proof}

Finally, we consider a connected component $C$ of $G - F$ with $\deg_F(C) = 1$.
We apply \Cref{branching:one_cc}, after which one of \Cref{branching:rpseudo_forest_standard,branching:naive,branching:connected} becomes applicable.

\begin{branching}
    \label{branching:one_cc}
    Let $C$ be a connected component of $G - F$ with $\deg_F(C) = 1$,
    and let $v$ be the unique vertex in $C$ with $\deg_F(v) = 1$.
    Let $P = (v_1, v_2, \ldots, v_t)$ be a shortest path in $C$ from $v = v_1$ to a vertex $v_t$ that lies on a cycle in $C$.
    The branching generates instances, defined as follows:
    \begin{align*}
    &I_i =
    (G - v_i, k - 1, F \cup \{v_1, \ldots, v_{i-1}\}, L|_{G - v_i},\mathcal{C}\cup \{L(v_i)\})
    \quad \text{for each } i \in [t], \text{ and}\\
    &I_{t+1} = (G, k, F \cup \{v_1, \ldots, v_t\}, L,\mathcal{C}).
    \end{align*}
\end{branching}

\begin{lemma}
\label{lem:one_neighb}
    In \Cref{branching:one_cc}, the path $P$ contains at most $2r$ vertices.
\end{lemma}

\begin{proof}
    Let $v$ be the unique vertex of $C$ adjacent to $F$.
    If $v$ lies on a cycle of $G[C]$, then $|V(P)|=1$.
    
    Let $Z$ be the set of vertices lying on cycles of $G[C]$.
    Let $X$ be the vertex set of the component of $G[C]-Z$ containing $v$,
    and let $T\coloneq G[X]$.
    Root $T$ at $v$, and attach a new leaf for each edge between $X$ and $Z$.
    Let $T'$ be the resulting tree.

    Every leaf of $T'$ is a new vertex, since otherwise a vertex of $G-F$ would have degree at most one in $G$.
    Moreover, each component of $G[Z]$ contains a cycle, and no component of $G[Z]$ has more than one edge to $X$.
    Hence, $T'$ has at most $r$ leaves.

    Every vertex of $T'$ with exactly one child has degree two in $G$.
    Thus, no two such vertices are adjacent.


    Consider a root-to-leaf path $Q$ in $T'$.
    Let $a$ and $b$ be the numbers of vertices of $Q$ with exactly one child and at least two children, respectively.
    Since $T'$ has at most $r$ leaves, we have $b\le r-1$.
    Moreover, $a\le b+1$, since no two vertices with exactly one child are adjacent.
    Therefore, $|V(Q)|=a+b+1\le 2b+2\le 2r$.
    Since each new leaf corresponds to a vertex on a cycle of $G[C]$, we obtain $|V(P)|\le 2r$.
\end{proof}

\subsection{Running Time Analysis}

To analyze the running time of \Search, we define a measure for an input instance $I = (G, k, F, L, \mathcal{C})$ as
\begin{align*}
    \mu(I) = k + \alpha \cdot \bigl(r \cdot \contcomp{G[F]} - \excess{G[F]}\bigr),
\end{align*}
where $\alpha \geq 1$ is a constant to be specified later.
This measure has been used previously in the design of FPT algorithms for \prb{$r$-Pseudoforest Deletion Set}~\cite{rpseudoforest:Tsur25}.
Note that none of the reduction rules increases $\mu(I)$.

\medskip
\noindent
\textbf{\Cref{branching:rpseudo_forest_standard}.}
By the same reasoning as in the analysis of \Cref{branching:lforest_standard}
in \Cref{sec:lforest}, we only consider the case $|U| = 3$ and let $U = \{u_1,u_2,u_3\} \subseteq V(G)\setminus F$ be a connected vertex set satisfying the conditions of \Cref{branching:rpseudo_forest_standard}.
This branching rule produces the following instances:
\begin{align*}
    &I_i = (G - u_i, k - 1, F \cup U_{i-1}, L|_{G - u_i},\mathcal{C}\cup \{L(u_i)\}) ~\text{for each $i \in [3]$, and} &\\
    &I_{4} = (G, k,F \cup U, L,\mathcal{C}),&
\end{align*}

In the instances $I_1,I_2$, and $I_3$, the parameter $k$ decreases by one, and none of the remaining terms in the measure increases.
Thus, $\mu(I_i) \leq \mu(I) - 1$ for each $i \in [3]$.

In the instance $I_4$, the parameter $k$ remains unchanged.
Since $\deg_F(U) \geq 2$ and $G[U]$ is connected, $\contcomp{G[F]}$ does not increase.
If $\contcomp{G[F]}$ decreases, then the measure decreases by at least $\alpha r$.
Otherwise, $\contcomp{G[F]}$ remains unchanged and $\excess{G[F]}$ increases by at least one, which decreases the measure by at least $\alpha$.
In either case, $\mu(I_4) \leq \mu(I) - \alpha$ since $r \geq 1$.
Therefore, the branching vector for \Cref{branching:rpseudo_forest_standard} is $(1, 1, 1, \alpha)$.

\medskip
\noindent
\textbf{\Cref{branching:shortest_path}.}
Let $P = (v_1, v_2, \ldots, v_t)$ be a path satisfying the conditions of
\Cref{branching:shortest_path}.
This branching rule generates the following instances:
\begin{align*}
    &I_i =
    (G - v_i, k - 1, F \cup \{v_1, \ldots, v_{i-1}\}, L|_{G - v_i},\mathcal{C}\cup \{L(v_i)\})
    \quad \text{for each } i \in [t], \text{ and}\\
    &I_{t+1} = (G, k, F \cup \{v_1, \ldots, v_t\}, L,\mathcal{C}).
\end{align*}

For each $i \in [t]$, the parameter $k$ decreases by one in $I_i$, and none of the other terms in the measure increases.
Hence, $\mu(I_i) \leq \mu(I) - 1$ for every $i \in [t]$.

In the instance $I_{t+1}$, the parameter $k$ remains unchanged and $\excess{G[F]}$ does not decrease.
By the choice of $P$, the number of connected components $\contcomp{G[F]}$ strictly decreases.
Therefore, $\mu(I_{t+1}) \leq \mu(I) - \alpha r$.

Therefore, the branching vector for \Cref{branching:shortest_path} is
$(\underbrace{1, \ldots, 1}_{t \text{ times}},\, \alpha r)$, where $t \leq 8r^2+12r+5$ by \Cref{obs:output_rank}.

\medskip
\noindent
\textbf{\Cref{branching:naive}.}
Let $C$ be a connected component of $G - F$ satisfying the conditions of \Cref{branching:naive}, and let $C'$ be the set obtained by applying \Cref{alg:cont_set} to $C$.
This branching rule produces the following instances:
\begin{align*}
    &I_i =
    (G - v_i, k - 1, F \cup \{v_1, \ldots, v_{i-1}\}, L|_{G - v_i},\mathcal{C}\cup \{L(v_i)\})
    \quad \text{for each } i \in [|C'|], \text{ and}\\
    &I_{|C'|+1} = (G, k, F\cup C', L,\mathcal{C}).
\end{align*}

For each $i \in [|C'|]$, the parameter $k$ decreases by one in $I_i$, and none of the other terms in the measure increases.
Hence, $\mu(I_i) \leq \mu(I) - 1$ for every $i \in [|C'|]$.

In the instance $I_{|C'|+1}$, the parameter $k$ remains unchanged.
Since $\deg_F(C') \geq 2$, the graph $G[F \cup C']$ contains a cycle that includes at least one vertex of $C'$.
Thus, by \Cref{fds:subgraph}, $\excess{G[F \cup C']}$ is greater than $\excess{G[F]}$ by at least one.
Moreover, since $G[C']$ is connected, $\contcomp{G[F]}$ remains unchanged.
Therefore, $\mu(I_{|C'|+1}) \leq \mu(I) - \alpha$.

By \Cref{lem:output_size} and the assumption of \Cref{branching:naive}, we have $|C'|\le(4r+3)r+(r+1)=4r^2+4r+1$.
Therefore, the branching vector for \Cref{branching:naive} is
$(\underbrace{1,\ldots,1}_{|C'|\text{ times}},\,\alpha)$, where $|C'|\le4r^2+4r+1$.

\medskip
\noindent
\textbf{\Cref{branching:connected}.}
Let $C$ be a connected component of $G - F$ satisfying the conditions of \Cref{branching:connected}, and let $C'$ be the set obtained by applying \Cref{alg:cont_set} to $C$.
This branching rule generates the following instances:
\begin{align*}
    I_i =
    (G - v_i, k - 1, F \cup \{v_1, \ldots, v_{i-1}\},L|_{G - v_i},\mathcal{C}\cup \{L(v_i)\})
    \quad \text{for each } i \in [|C'|].
\end{align*}

For each $i \in [|C'|]$, the parameter $k$ decreases by one, and none of the other terms in the measure increases.
Hence, $\mu(I_i) \leq \mu(I) - 1$ for every $i \in [|C'|]$.

By \Cref{obs:output_rank}, we have $|C'| \leq 8r^2+12r+5$.
Therefore, the branching vector for \Cref{branching:connected} is
$(\underbrace{1, \ldots, 1}_{|C'| \text{ times}})$, where $|C'| \leq 8r^2+12r+5$.

\medskip
\noindent
\textbf{\Cref{branching:one_cc}.}
Let $P = (v_1, v_2, \ldots, v_t)$ be a path satisfying the conditions of \Cref{branching:one_cc}.
This branching rule produces the following instances:
\begin{align*}
    &I_i =
    (G - v_i, k - 1, F \cup \{v_1, \ldots, v_{i-1}\}, L|_{G - v_i},\mathcal{C}\cup \{L(v_i)\})
    \quad \text{for each } i \in [t], \text{ and}\\
    &I_{t+1} = (G, k, F \cup \{v_1, \ldots, v_t\}, L,\mathcal{C}).
\end{align*}

For each $i \in [t]$, the parameter $k$ decreases by one, and none of the other terms in the measure increases.
Hence, $\mu(I_i) \leq \mu(I) - 1$ for every $i \in [t]$.

In the instance $I_{t+1}$, the measure may not decrease.
If $\excess{G[F \cup \{v_1, \ldots, v_t\}]} > \excess{G[F]}$ (in particular, if $v_t$ has a loop), then the measure decreases by at least $\alpha$.
Otherwise, after applying this branch, at least one of
\Cref{branching:rpseudo_forest_standard,branching:connected,branching:shortest_path,branching:naive}
becomes applicable, even after applying reduction rules, since the vertex $v_t$ lies on a non-loop cycle and has two incident edges to same connected component of $G - (F \cup \{v_1, \ldots, v_t\})$.

Since $\alpha = 1$, the worst-case continuation is given by
\Cref{branching:shortest_path}, which has at most $8r^2+12r+6$ branches, each decreasing the measure by at least one.
By \Cref{lem:one_neighb}, we have $t \leq 2 r$.
Therefore, the branching vector for \Cref{branching:one_cc} is 
$(1, 1, \ldots, 1)$, where the value $1$ appears at most $2r+(8r^2+12r+6)=8r^2+14r+6$ times.

\medskip
\noindent
\textbf{Running Time and Correctness.}

We analyze the running time of \ComputingProductList.
Let
$I = (\hat{G} - S, \hat{k} - |S|, \hat{S} \setminus S, L,\{ \{v\} \mid v \in S\})$
be an instance of \Search arising during the execution of \ComputingProductList.
For this instance,
\[
    \mu(I) \leq (\hat{k} - |S|) + \alpha \cdot r (\hat{k} - |S|)
    = (\hat{k} - |S|)(\alpha \cdot r + 1).
\]

Assume $r \geq 1$ and set $\alpha = 1$\rev{, thus $\alpha r \geq 1$}.
All (possibly combined) branching steps decrease the measure by at least one and have at most $R+1$ branches, where $R\coloneq8r^2+14r+5$.
\rev{Since $\alpha r \geq 1$, the running time of \Search on instance $I$ is bounded by $(R + 1)^{\mu(I)} \cdot n^{O(1)}$.}
Substituting the bound on $\mu(I)$ yields $(R + 1)^{(\hat{k} - |S|)(1 + \alpha r)} \cdot n^{O(1)}$.

Since the algorithm enumerates all subsets $S \subseteq \hat{S}$, the total running time of \ComputingProductList is
\begin{align*}
    \sum_{S \subseteq \hat{S}} \rev{(R + 1)}^{(\hat{k} - |S|)(1 + \alpha r)} \cdot n^{O(1)}
    \leq \sum_{i=0}^{\hat{k}} \binom{\hat{k}}{i} \rev{(R + 1)}^{i(1 + r)} \cdot n^{O(1)} = (1 + (R + 1)^{1 + r})^{\hat{k}} \cdot n^{O(1)}.
\end{align*}

By \Cref{lem:comp_forb}, \Cref{branching:connected} is safe.
The correctness follows from the safeness of all reduction and branching rules and from the fact that all cases are exhaustively covered.

\begin{theorem}
\label{thm:rpseudoforest_comp_rep}
Let $(G, k)$ be an instance of \prb{$r$-Pseudoforest Deletion Set}.
There exists an algorithm that computes a $k$-product list for $r$-pseudoforest deletion sets of $G$ in FPT time parameterized by $r + k$.
\end{theorem}

\section{Applications}
\label{sec:application}
In this section, we show that $k$-product lists yield FPT algorithms for several variants of \delprb, including the weighted variant (\Cref{subsec:weighted}), the nonpartisan variant (\Cref{subsec:fair}), the hereditary-constraint variant (\Cref{subsec:hereditary}), the reconfiguration variant (\Cref{subsec:reconfiguration}), the connected variant (\Cref{subsec:connected}), the conflict-free variant (\Cref{subsec:conflict}), the simultaneous variant (\Cref{subsec:simul}), and the diverse variant (\Cref{subsec:diverse}).

\subsection{Weighted \delprb\ with Cardinality Constraints}
\label{subsec:weighted}

We first consider a weighted variant of \delprb\ with an explicit cardinality constraint, defined as follows.

\begin{center}
	\fbox{
		\parbox{0.9\hsize}{
			\prb{Weighted {\delprb}}
			\begin{listing}{\textbf{Input:}}
				\item[\textbf{Input:}] A vertex-weighted graph $G$ with a weight function $w \colon V(G) \to \mathbb{Z}$ and a positive integer $k$.
				\item[\textbf{Task:}] Find a minimum-weight $\mathcal{F}$-deletion set of size at most $k$.
			\end{listing}
		}
	}
\end{center}
The weight function is allowed to take negative values. 
An $\mathcal{F}$-deletion set minimizing the total weight is called an \emph{optimal} deletion set.

The \prb{Weighted Feedback Vertex Set} problem has been studied from the perspective of parameterized complexity. 
Becker et al.~\cite{WeightedFVS:BeckerBG00} presented a randomized FPT algorithm running in $O(6^k \cdot kn^2)$ time for the case of positive vertex weights. 
Chen et al.~\cite{WeightedFVS:ChenFLLV08} subsequently obtained a deterministic algorithm with running time $O(5^k \cdot kn^2)$. 
Later, Agrawal et al.~\cite{weighted:AgrawalKLS16} further improved the running time to $O^*(3.618^k)$ while studying \prb{Block Graph Vertex Deletion}.

We establish fixed-parameter tractability for the weighted variant of \delprb.

\begin{theorem}
	\label{thm:weighted}
    Let $\mathcal{F}$ be a class of graphs closed under taking subgraphs.
    Suppose that there exists an algorithm that computes a $k$-product list for $\mathcal{F}$-deletion sets in FPT time with respect to $k$.
    Then, \prb{Weighted {\delprb}} is fixed-parameter tractable when parameterized by $k$.
\end{theorem}


We first observe that the case of strictly positive vertex weights is straightforward.
If $w(v) > 0$ for all $v \in V(G)$, then every optimal solution to
\prb{Weighted $\mathcal{F}$-Deletion Set} is inclusion-wise minimal.
Indeed, adding any redundant vertex strictly increases the total weight.
Consequently, it suffices to optimize over the minimal $\mathcal{F}$-deletion sets of size at most $k$.
A $k$-product list handles them implicitly: for each product representation $\mathcal{C} \in \Lclass$, one selects a minimum-weight vertex from each member of $\mathcal{C}$.

We now turn to the general case, where vertex weights may be arbitrary integers, including negative values.
In this setting, an optimal solution may contain vertices that are not in a minimal $\mathcal{F}$-deletion set in order to decrease the total weight.
We therefore need to handle such vertices.

The following observation captures the structure of optimal solutions.

\begin{observation}
	\label{obs:weighted_minimal}
    Let $\mathcal{F}$ be a class of graphs closed under taking subgraphs, and let $(G,w,k)$ be an instance of
	\prb{Weighted \delprb}.
	There exists an optimal solution $S$ of the form $S=S'\cup N$, where $S'$ is a minimal $\mathcal{F}$-deletion set and every vertex in $N\subseteq V(G)\setminus S'$ has negative weight.
\end{observation}

We now describe an algorithm for \prb{Weighted \delprb}.
Let $\Lclass$ be a $k$-product list for $\mathcal{F}$-deletion sets of $G$.
For each $k$-product representation $\mathcal{C} = \{C_1, C_2,\ldots,C_q\} \in \Lclass$ with $q \le k$, we construct a candidate solution as follows.
For each $i \in [q]$, let $v_i \in C_i$ be a vertex of minimum weight in $C_i$, that is, $w(v_i) = \min_{v \in C_i} w(v)$, and define $S' \coloneqq \{v_1, v_2, \ldots, v_q\}$.
By construction, $S'$ has minimum weight among all sets obtained by selecting exactly one vertex from each member of $\mathcal{C}$.
If $|S'| < k$, we extend $S'$ by adding vertices from $V(G)\setminus S'$ with the smallest negative weights, as long as the cardinality constraint $|S| \leq k$ is satisfied.
Formally, let $N$ consist of up to $(k-q)$ vertices of minimum weight in $V(G)\setminus S'$ with $w(v) < 0$, and define $S \coloneqq S' \cup N$.
Among all solutions constructed in this manner, we return one with the minimum total weight.
The correctness of this procedure follows directly from \Cref{obs:weighted_minimal}, which completes the proof of \Cref{thm:weighted}.

As a corollary, we obtain the following result.

\begin{corollary}
	\label{cor:weighted}
	\prb{Weighted $\ell$-Forest Deletion Set} and
	\prb{Weighted $r$-Pseudoforest Deletion Set}
	are fixed-parameter tractable when parameterized by $k+\ell$ and $k+r$, respectively.
\end{corollary}

\subsection{Nonpartisan Variant}
\label{subsec:fair}

Various notions of fairness have been studied for solutions to combinatorial optimization problems.
Based on a fairness measure for shortest paths introduced by Bentert et al.~\cite{fair:BentertKN23},
Purohit et al.~\cite{fair:PurohitSSU25} studied the following nonpartisan variant of the \prb{Feedback Vertex Set} problem.\footnote{Purohit et al.~\cite{fair:PurohitSSU25} refer to this problem as \prb{Fair Feedback Vertex Set}.
We use the term \emph{Nonpartisan} to distinguish it from other fair variants (e.g.,~\cite{fair:KaneshMMS21}).}

\begin{center}
	\fbox{
		\parbox{0.9\hsize}{
			\prb{Nonpartisan {\delprb}}
			\begin{listing}{\textbf{Input:}}
				\item[\textbf{Input:}]
				A graph $G$ whose vertex set is partitioned into
				$\mathcal{V} = (V_1, V_2, \ldots, V_p)$
				(with $V_i \neq \emptyset$ for all $i \in [p]$),
				a vector $c = (c_1, c_2, \ldots, c_p) \in \mathbb{Z}_{\ge 0}^p$,
				and a positive integer $k$.
				\item[\textbf{Task:}]
				Find an $\mathcal{F}$-deletion set $S$ of size at most $k$
				such that $|S \cap V_i| \le c_i$ holds for every $i \in [p]$.
			\end{listing}
		}
	}
\end{center}

Purohit et al.~\cite{fair:PurohitSSU25} presented an $O^*(17^k)$-time fixed-parameter algorithm for the nonpartisan version of \prb{Feedback Vertex Set}.

In this subsection, we prove fixed-parameter tractability for the nonpartisan variant of \delprb.

\begin{theorem}
    \label{thm:fair}
    Suppose that there exists an algorithm that computes a $k$-product list for $\mathcal{F}$-deletion sets in FPT time with respect to $k$.
    Then, \prb{Nonpartisan {\delprb}} is fixed-parameter tractable when parameterized by $k$.
\end{theorem}

We now present a framework for solving \prb{Nonpartisan \delprb} based on product representations.
To this end, we rely on a polynomial-time algorithm for \prb{Maximum-Weight $b$-Matching}.
Let $G$ be a graph and let $b \colon V(G) \to \mathbb{Z}_{\ge 0}$ be a vertex-capacity function.
A \emph{$b$-matching} of $G$ is a function $f \colon E(G) \to \{0,1\}$ such that $\sum_{e \in \partial(v)} f(e) \le b(v)$ for all $v \in V(G)$,
where $\partial(v)$ denotes the set of edges incident to $v$.
For a vertex subset $S \subseteq V(G)$,
a $b$-matching $f$ is \emph{$S$-perfect} if $\sum_{e \in \partial(v)} f(e) = b(v)$ holds for all $v \in S$.
It is well known that deciding the existence of an $S$-perfect $b$-matching,
as well as finding one of maximum weight,
can be done in polynomial time~\cite{matching:Muller-HannemannS00}.

Let $(G, \mathcal{V}, c, k)$ be an instance of \prb{Nonpartisan $\mathcal{F}$-Deletion Set},
and let $\Lclass$ be a $k$-product list for $\mathcal{F}$-deletion sets of $G$.
For each product representation $\mathcal{C} \in \Lclass$,
we construct a bipartite graph $G_{\mathcal{C}}$ as follows.
The vertex set of $G_{\mathcal{C}}$ is $[p] \cup \mathcal{C}$, with bipartition $([p], \mathcal{C})$.
A vertex $i \in [p]$ is adjacent to a set $C \in \mathcal{C}$ if and only if $V_i \cap C \neq \emptyset$.
We define a vertex-capacity function $b_{\mathcal{C}}$ by setting $b_{\mathcal{C}}(i) = c_i$ for each $i \in [p]$ and $b_{\mathcal{C}}(C) = 1$ for each $C \in \mathcal{C}$.

We then check whether $(G_{\mathcal{C}}, b_{\mathcal{C}})$ admits a $\mathcal{C}$-perfect $b_{\mathcal{C}}$-matching.
Since $|V(G_{\mathcal{C}})| = p + |\mathcal{C}| \le p + k \le 2n$,
this check can be performed in polynomial time.
If such a $b_{\mathcal{C}}$-matching exists for some $\mathcal{C} \in \Lclass$,
the algorithm returns \ttyes; otherwise, it returns \ttno.

If the algorithm returns \ttyes,
we can construct a solution for \prb{Nonpartisan $\mathcal{F}$-Deletion Set} as follows.
Let $\mathcal{C} \in \Lclass$ be a product representation for which a $\mathcal{C}$-perfect $b_{\mathcal{C}}$-matching exists. 
For each $C \in \mathcal{C}$, the matching assigns $C$ to exactly one adjacent vertex $i \in [p]$.
We then select an arbitrary vertex $v \in C \cap V_i$ and include it in the solution.
Note that the number of selected vertices from each $V_i$ is bounded by $c_i$ due to the capacity constraints of $b_{\mathcal{C}}$.
Since $\mathcal{C}$ is a $k$-product representation for $\mathcal{F}$-deletion sets of $G$,
the resulting set is indeed an $\mathcal{F}$-deletion set.

For correctness, observe that every inclusion-minimal feasible solution $S$ to \prb{Nonpartisan $\mathcal{F}$-Deletion Set} is a minimal $\mathcal{F}$-deletion set and is therefore represented by some $\mathcal{C} \in \Lclass$.
Moreover, given such a solution $S$ and a product representation $\mathcal{C}$ that represents $S$,
one can construct a corresponding $\mathcal{C}$-perfect $b_{\mathcal{C}}$-matching by assigning each $C \in \mathcal{C}$ to an index $i \in [p]$ such that $S \cap C \subseteq V_i$.
Therefore, we complete the proof of \Cref{thm:fair}.

As a corollary, we obtain the following result.

\begin{corollary}
	\label{cor:fair}
	\prb{Nonpartisan $\ell$-Forest Deletion Set} and
	\prb{Nonpartisan $r$-Pseudoforest Deletion Set}
	are fixed-parameter tractable when parameterized by $k + \ell$ and $k + r$, respectively.
\end{corollary}

\subsection{Hereditary Property Constraint}
\label{subsec:hereditary}

We consider a \emph{graph property} $\Pi$, that is, a class of graphs closed under isomorphism.
Formally, if two graphs $G$ and $H$ are isomorphic and $G$ satisfies $\Pi$, then $H$ also satisfies $\Pi$.
A graph property $\Pi$ is said to be \emph{hereditary} if, for every graph $G$ satisfying $\Pi$, every induced subgraph of $G$ also satisfies $\Pi$.
Many natural graph classes are hereditary, such as edgeless graphs and minor-closed classes.

Tamura et al.~\cite{hereditary:TamuraI020} studied the problem of finding a feedback vertex set that additionally satisfies a given hereditary property.
In this subsection, we extend their framework from feedback vertex sets to general $\mathcal{F}$-deletion sets.

\begin{center}
	\fbox{
		\parbox{0.9\hsize}{
			\prb{$\Pi$-\delprb}
			\begin{listing}{}
				\item[\textbf{Input:}]
				A graph $G$ and a positive integer $k$.
				\item[\textbf{Task:}]
				Find an $\mathcal{F}$-deletion set
				$S \subseteq V(G)$ of size at most $k$
				such that the induced subgraph $G[S]$ satisfies $\Pi$.
			\end{listing}
		}
	}
\end{center}

For example, when $\Pi$ is the class of edgeless graphs, the corresponding problem for feedback vertex sets is known as \prb{Independent Feedback Vertex Set}.
This problem has been extensively studied in the context of parameterized complexity~\cite{IndepFVS:param_AgrawalGSS16,IndepFVS:param_LiP20,IndepFVS:param_MisraPRS12,IndepFVS:param_Song14,IndepFVS:param_TamuraIZ15}.

Tamura et al.~\cite{hereditary:TamuraI020} designed a fixed-parameter tractable algorithm for \prb{$\Pi$-Feedback Vertex Set} on unweighted graphs, based on the compact representations for feedback vertex sets shown by Guo et al.~\cite{CompactRep:GuoGHNW06}.
Their algorithm, however, crucially exploits structural properties specific to Guo et al.'s construction.
In order to establish fixed-parameter tractability for \prb{$\rmpi$-$\ell$-Forest Deletion Set} and
\prb{$\rmpi$-$r$-Pseudoforest Deletion Set},
we rely on structural properties of $k$-product lists of \Cref{thm:lforest_comp_rep,thm:rpseudoforest_comp_rep}.

Let $G$ be a graph on $n$ vertices.
Throughout this subsection, we assume that there exists an algorithm that decides whether a given graph $H$ on $h$ vertices satisfies $\Pi$ in time at most $t(h)$.
In addition, let $\mathcal{F}$ be the class of $\ell$-forests or $r$-pseudoforests.

\begin{theorem}
	\label{thm:hereditary}
	Let $\Pi$ be a hereditary graph property such that there exists an algorithm that decides whether a given graph $H$ on $h$ vertices satisfies $\Pi$ in time at most $t(h)$.
	Then, \prb{$\Pi$-$\ell$-forest Deletion Set} and \prb{$\Pi$-$r$-pseudoforest Deletion Set}
	are fixed-parameter tractable when parameterized by $k + \ell$ and $k + r$, respectively.
\end{theorem}

We begin by observing a basic structural property of our $k$-product lists in \Cref{sec:lforest,sec:pseudoforest}.

\begin{observation}
	\label{obs:product_connected}
	Let $\Lclass$ be a $k$-product list for $\mathcal{F}$-deletion sets in a graph $G$ obtained as described in \Cref{sec:lforest,sec:pseudoforest}.
	For every $\mathcal{C} \in \Lclass$ and every $C \in \mathcal{C}$ with $|C|\ge 2$, the induced subgraph $G[C]$ is a path.    
\end{observation}

Indeed, a label with at least two vertices is created only by \Cref{red:deg2_lforest} or \Cref{red:deg2_rforest}, which, by induction, joins two paths at their endpoints by a single edge, since $u_1$ and $u_2$ have degree two and $u_1u_2$ is the unique edge between them.

In what follows, we describe a preprocessing procedure applied to a single $k$-product representation $\mathcal{C}$ for $\ell$-forest or $r$-pseudoforest deletion sets.
A $k$-product representation $\mathcal{C}$ is said to be \emph{good} if there is a set $S$ obtained by selecting exactly one vertex from each member of $\mathcal C$ such that $G[S]$ satisfies $\Pi$.
If every set in $\mathcal{C}$ has size at most $\alpha$, then one can test whether $\mathcal{C}$ is good by naive brute force.

\begin{observation}
    \label{obs:good_test}
	Let $G$ be a graph.
	Given a $k$-product representation $\mathcal{C}$ of $\mathcal{F}$-deletion sets in $G$, assume that $|C| \le \alpha$ for every $C \in \mathcal{C}$.
	Then one can determine whether $\mathcal{C}$ is good in time $O(\alpha^k \cdot t(k)) \cdot n^{O(1)}$.
\end{observation}

We call a set $C \in \mathcal{C}$ a \emph{singleton} if $|C| = 1$, and a \emph{multiple set} otherwise.
We next bound the number of edges between two multiple sets.

\begin{lemma}
	\label{lem:comprep_edges_forest}
	Let $G$ be a graph and let $\mathcal{C}$ be a $k$-product representation in a $k$-product list for $\ell$-forest deletion sets produced by the algorithm of \Cref{sec:lforest}.
	For any two distinct multiple sets $C, C' \in \mathcal{C}$,
    the number of edges between $C$ and $C'$ is at most $4(\ell+3)$.
\end{lemma}

\begin{proof}   
    Let $u \in C$ and $u' \in C'$ be arbitrary vertices, and define $G' \coloneq G[(C \setminus \{u\}) \cup (C' \setminus \{u'\})]$.

    We first show that $G'$ is an $\ell$-forest.
    Since $\mathcal{C}$ is a $k$-product representation, selecting $u$ from $C$, $u'$ from $C'$, and one arbitrary vertex from every other member yields an $\ell$-forest deletion set $S$. 
    Thus, $G'$ is an induced subgraph of $G-S$ and therefore an $\ell$-forest, since the class of $\ell$-forests is closed under taking subgraphs.
    
    By \Cref{obs:product_connected}, both $G[C]$ and $G[C']$ are paths.
    Thus, each of $G[C\setminus\{u\}]$ and $G[C'\setminus\{u'\}]$ has at most two connected components.
    Thus, $G'$ consists of at most four connected components.

    By \Cref{obs:product_connected}, both $G[C]$ and $G[C']$ are paths.
    Each of $G[C \setminus \{u\}]$ and $G[C' \setminus \{u'\}]$ has at most two connected components.
    Contracting each of these at most four components to a single vertex yields a multigraph $H$ on at most four vertices; no edge between $C$ and $C'$ is contracted.
    Since the class of $\ell$-forests is closed under edge contraction, $H$ is also an $\ell$-forest and thus has at most $\ell + 4 - 1 = \ell + 3$ edges.
    Therefore, the number of edges between $C \setminus \{u\}$ and $C' \setminus \{u'\}$ is at most $\ell + 3$.

    Let $a=|C|$, $b=|C'|$, and let $m$ be the number of edges between $C$ and $C'$. 
    Summing over all $ab$ choices of $u\in C$ and $u'\in C'$, each of these $m$ edges is counted exactly $(a-1)(b-1)$ times. 
    Since every choice leaves at most $\ell+3$ edges between the two sets, we have $(a-1)(b-1)m\le ab(\ell+3)$. 
    As $a,b\ge2$, this implies $m\le4(\ell+3)$.
\end{proof}

\begin{lemma}
	\label{lem:comprep_edges_pseudoforest}
	Let $G$ be a graph and let $\mathcal{C}$ be a $k$-product 
    representation in a $k$-product list produced by the algorithm of \Cref{sec:pseudoforest}.
	For any two distinct multiple sets $C, C' \in \mathcal{C}$, 
    the number of edges between $C$ and $C'$ is at most $4(2r+2)$.
\end{lemma}

\begin{proof}
    Let $u \in C$ and $u' \in C'$ be arbitrary vertices, and define $G' \coloneq G[(C \setminus \{u\}) \cup (C' \setminus \{u'\})]$.

    We first show that $G'$ is an $r$-pseudoforest.
    Since $\mathcal{C}$ is a $k$-product representation, selecting $u$ from $C$, $u'$ from $C'$, and one arbitrary vertex from every other member yields an $r$-pseudoforest deletion set $S$. 
    Thus, $G'$ is an induced subgraph of $G-S$ and therefore an $r$-pseudoforest by \Cref{obs:rpseudo_subgraph}.
    
    By \Cref{obs:product_connected}, both $G[C]$ and $G[C']$ are paths.
    Thus, each of $G[C\setminus\{u\}]$ and $G[C'\setminus\{u'\}]$ has at most two connected components.
    Thus, $G'$ consists of at most four connected components.

    Contract each connected component of $G[C\setminus\{u\}]$ and $G[C'\setminus\{u'\}]$ separately. 
    The resulting multigraph $H$ is bipartite with at most four vertices and is an $r$-pseudoforest. 
    Let $c$ be the number of components of $H$ containing an edge. 
    Then $c\le2$, and, for $c\ge1$, these components have at most $4-c+cr\le2r+2$ edges, since $r\ge1$. 
    The case $c=0$ is immediate. 
    No edge between the two sets is contracted, giving the required bound.

    Let $a=|C|$, $b=|C'|$, and let $m$ be the number of edges between $C$ and $C'$. 
    Summing over all $ab$ choices of $u\in C$ and $u'\in C'$, each of these $m$ edges is counted exactly $(a-1)(b-1)$ times. 
    Since every choice leaves at most $2r+2$ edges between the two sets, we have $(a-1)(b-1)m\le ab(2r+2)$. 
    As $a,b\ge2$, this implies $m\le4(2r+2)$.
\end{proof}

Let $\singletons{\mathcal{C}}$ denote the set of vertices that appear in singleton sets of $\mathcal{C}$.
We introduce a modified version of the reduction rule of Tamura et al.~\cite{hereditary:TamuraI020}.
For a multiple set $C \in \mathcal{C}$ and a subset $X \subseteq \singletons{\mathcal{C}}$, define
\begin{align*}
	C_{X} \coloneq
	\{u \in C \mid N_{\singletons{\mathcal{C}}}(u) = X
	\text{ and $u$ has no neighbor in any multiple set $C' \in \mathcal{C} \setminus \{C\}$}\}.
\end{align*}

\begin{redrule}
	\label{rule:hereditary_kernel}
    If there exists a multiple set $C \in \mathcal{C}$ and a subset $X \subseteq \singletons{\mathcal{C}}$ such that $|C_X| \ge 2$, 
    then delete all but one vertex from $C_X$.
\end{redrule}

The reduction rule can be applied in polynomial time.
Moreover, the resulting family remains a $k$-product representation, 
since removing vertices from sets in $\mathcal{C}$ preserves the defining property of product representations.

\begin{lemma}
	\label{lem:hereditary_size}
    Let $\mathcal{C}$ be a $k$-product representation in a $k$-product list for $\ell$-forest deletion sets (resp., $r$-pseudoforest deletion sets) in $G$ produced by the algorithm of \Cref{sec:lforest} (resp., \Cref{sec:pseudoforest}).
    Let $\mathcal{C}^*$ be the $k$-product representation obtained by exhaustively applying \Cref{rule:hereditary_kernel} to $\mathcal{C}$.
	Then the following hold:
	\begin{enumerate}
        \item Each set in $\mathcal{C}^*$ has size at most $2^k + 4(\ell+3)\binom{k}{2}$ in the $\ell$-forest case, and at most $2^k+4(2r+2)\binom{k}{2}$ in the $r$-pseudoforest case.\label{lem:hereditary_size_first}
		\item $\mathcal{C}$ is good if and only if $\mathcal{C}^*$ is good. \label{lem:hereditary_size_second}
	\end{enumerate}
\end{lemma}

\begin{proof}
    We first prove the size bound in Item~\ref{lem:hereditary_size_first}.
    Consider first the case in which $\mathcal{F}$ is the class of $\ell$-forests.
    Suppose that $\mathcal{C}$ contains a multiple set $C$ with more than $2^k + 4(\ell+3)\cdot \binom{k}{2}$ vertices.
    Since $|\singletons{\mathcal{C}}| \le k$, there are at most $2^k$ distinct subsets of $\singletons{\mathcal{C}}$.
    By \Cref{lem:comprep_edges_forest}, the number of vertices in $C$ that have neighbors in other multiple sets of $\mathcal{C}$ is at most $4(\ell+3)\cdot \binom{k}{2}$ (this also holds after applying \Cref{rule:hereditary_kernel}, since the rule only removes vertices).
    Hence, there exists a subset $X \subseteq \singletons{\mathcal{C}}$ such that $|C_X| \ge 2$, and \Cref{rule:hereditary_kernel} is applicable.
	Therefore, after exhaustive application of the rule, every set in $\mathcal{C}^*$ has size at most $2^k + 4(\ell+3)\cdot \binom{k}{2}$.
	The argument for $r$-pseudoforests follows analogously using \Cref{lem:comprep_edges_pseudoforest}, giving the bound $2^k+4(2r+2)\binom{k}{2}$.

    We next prove \Cref{lem:hereditary_size_second}.
    Let $\mathcal{C}'$ be the representation obtained from $\mathcal{C}$ by a single application of \Cref{rule:hereditary_kernel}.
    The reverse direction is immediate: if $\mathcal{C}'$ is good, then so is $\mathcal{C}$.

    For the converse, assume that $\mathcal{C}$ is good, and let $S$ be a set obtained by selecting exactly one vertex from each member of $\mathcal{C}$ such that $G[S]$ satisfies $\Pi$.
    Suppose that \Cref{rule:hereditary_kernel} is applied to a multiple set $C \in \mathcal{C}$ and a subset $X \subseteq \singletons{\mathcal{C}}$, and that all but one vertex $u \in C_X$ are removed.
    If $u \in S$, then $S$ is still represented by $\mathcal{C}'$, and hence $\mathcal{C}'$ is good.

    Otherwise, there exists a vertex $u' \in C_X \setminus \{u\}$ with $u' \in S$.
    Let $S' \coloneq (S \setminus \{u'\}) \cup \{u\}$. 
    Since $u$ and $u'$ belong to the same set $C$, the set $S'$ is represented by $\mathcal{C}'$.

    We now show that $G[S]$ and $G[S']$ are isomorphic.
    By definition of $C_X$, the vertices $u$ and $u'$ have identical neighborhoods in $\singletons{\mathcal{C}}$ and no neighbors in any other multiple set of $\mathcal{C}$.
    Furthermore, both $S$ and $S'$ contain exactly one vertex from every member of $\mathcal{C}$ and no other vertices.
    Consequently, the induced subgraphs $G[S]$ and $G[S']$ differ only by the exchange of $u$ and $u'$, and hence are isomorphic.
    Since $\Pi$ is hereditary and closed under isomorphism, and $G[S]$ satisfies $\Pi$, it follows that $G[S']$ also satisfies $\Pi$, which implies that $\mathcal{C}'$ is good.
\end{proof}

We are now ready to prove \Cref{thm:hereditary}.
We first compute a $k$-product list $\Lclass$ using the algorithms described in \Cref{sec:lforest,sec:pseudoforest}.
For each $\mathcal{C} \in \Lclass$, we exhaustively apply \Cref{rule:hereditary_kernel} and obtain a reduced representation $\mathcal{C}^*$.
By \Cref{lem:hereditary_size_first}, the size of each set in $\mathcal{C}^*$ is bounded by a function of $k+\ell$ (resp., $k+r$), and by \Cref{lem:hereditary_size_second}, this reduction preserves goodness.
Combined with \Cref{obs:good_test}, enumerating all sets obtained by selecting exactly one vertex from each member of $\mathcal{C}^*$ and checking whether the resulting induced graph satisfies $\Pi$ takes time $f(k+\ell)\cdot t(k)\cdot n^{O(1)}$ (resp., $f(k+r)\cdot t(k)\cdot n^{O(1)}$), 
where $f$ is some computable function.
Moreover, $(G,k)$ has a solution if and only if some $\mathcal{C} \in \Lclass$ is good: if $S$ is a solution, then a minimal $\mathcal{F}$-deletion set $S' \subseteq S$ is represented by some $\mathcal{C} \in \Lclass$ and $G[S']$ satisfies $\Pi$ since $\Pi$ is hereditary, and the converse is immediate.
This completes the proof of \Cref{thm:hereditary}.

\subsection{Reconfiguration Problems}
\label{subsec:reconfiguration}

The reconfiguration framework~\cite{reconf:ItoDHPSUU11} has been extensively studied for a wide range of combinatorial search problems; see the surveys~\cite{reconf:survey_BousquetMNS24,reconf:survey_Nishimura18,reconf:survey_Heuvel13}.
Two $\mathcal{F}$-deletion sets $S,S'$ of a graph $G$ are \emph{adjacent under the Token Jumping (TJ) rule} if there exist distinct vertices $u,v\in V(G)$ such that $S\setminus S'=\{u\}$ and $S'\setminus S=\{v\}$.
Two $\mathcal{F}$-deletion sets $S$ and $S'$ are \emph{reconfigurable under the TJ rule} if there exists a sequence $S_1,S_2,\ldots,S_p$ of $\mathcal{F}$-deletion sets such that $S_1=S$, $S_p=S'$, and $S_i$ and $S_{i+1}$ are adjacent under the TJ rule for every $i\in[p-1]$.
We consider the following problem.

\begin{center}
	\fbox{
		\parbox{0.9\hsize}{
			\prb{{\delprb} Reconfiguration}
			\begin{listing}{\textbf{Input:}}
				\item[\textbf{Input:}]
				An unweighted graph $G$, a positive integer $k$,
				and two $\mathcal{F}$-deletion sets
				$S^{\texttt{s}}$ and $S^{\texttt{t}}$ of size exactly $k$.
				\item[\textbf{Task:}]
				Decide whether $S^{\texttt{s}}$ can be reconfigured into $S^{\texttt{t}}$
				under the TJ rule.
			\end{listing}
		}
	}
\end{center}

It is known that \prb{Feedback Vertex Set Reconfiguration} is $\PSPACE$-complete~\cite{reconf:BousquetHKMS23,reconf:MouawadN0SS17}. 
Moreover, Mouawad et al.~\cite{reconf:MouawadN0SS17} presented a kernel with $O(k^3)$ vertices for this problem.
In this subsection, we investigate the fixed-parameter tractability of reconfiguration problems. 

\begin{theorem}
	\label{thm:reconfiguration}
    Let $\mathcal{F}$ be a class of graphs closed under taking subgraphs.
    Suppose that there exists an algorithm that computes a $k$-product list for $\mathcal{F}$-deletion sets in FPT time with respect to $k$.
    Then, \prb{{\delprb} Reconfiguration} is fixed-parameter tractable when parameterized by $k$.
\end{theorem}

We begin with the following lemma.

\begin{lemma}
\label{lem:reconf_same_node}
    Let $\mathcal{C}$ be a $k$-product representation for $\mathcal{F}$-deletion sets of $G$, and let $S,S'\subseteq V(G)$ be sets represented by $\mathcal{C}$ such that $|S|=|S'|$.
    Then $S$ and $S'$ are reconfigurable under the TJ rule.
\end{lemma}

\begin{proof}
    The case $S = S'$ is trivial.
    Assume that $S\neq S'$ and choose $x\in S\setminus S'$.
    If $S\setminus\{x\}$ intersects every member of $\mathcal{C}$, choose any $y\in S'\setminus S$.
    Then $(S\setminus\{x\})\cup\{y\}$ is represented by $\mathcal{C}$.

    Otherwise, there exists $C\in\mathcal{C}$ such that $S\cap C=\{x\}$. 
    Since $S'$ is represented by $\mathcal{C}$, choose $y\in S'\cap C$. 
    Hence $(S\setminus\{x\})\cup\{y\}$ is represented by $\mathcal{C}$.

    In either case, letting $T=(S\setminus\{x\})\cup\{y\}$, we have $|T|=|S|$ and $|T\setminus S'|=|S\setminus S'|-1$. 
    Since $\mathcal{F}$ is closed under taking subgraphs, every set represented by $\mathcal{C}$ is an $\mathcal{F}$-deletion set.
    Therefore, the set $T$ is an $\mathcal{F}$-deletion set.
    Repeating the above argument yields a reconfiguration sequence from $S$ to $S'$.
\end{proof}

Let $\Lclass$ be a $k$-product list for $\mathcal{F}$-deletion sets.
Given two $k$-product representations $\mathcal{C},\mathcal{C}'\in\Lclass$, let $\confgraph{\mathcal{C}}{\mathcal{C}'}$ be the bipartite graph with bipartition $(\mathcal{C},\mathcal{C}')$ in which $C\in\mathcal{C}$ and $C'\in\mathcal{C}'$ are adjacent if and only if $C\cap C'\neq\emptyset$.
Let $\mu(\mathcal{C},\mathcal{C}')$ denote the size of a maximum matching in $\confgraph{\mathcal{C}}{\mathcal{C}'}$.

The next lemma characterizes when two $k$-product representations represent sets with size $k$ that are equal or adjacent under the TJ rule.
Let $A \symdif B$ denote the symmetric difference of $A$ and $B$, defined by $A \symdif B = (A \setminus B) \cup (B \setminus A)$.

\begin{lemma}
    \label{lem:reconf_edge_test}
    Let $\mathcal{C},\mathcal{C}'\in\Lclass$, and assume that $|V(G)|\ge k$.
    There exist size-$k$ sets $A,B\subseteq V(G)$ represented by $\mathcal{C}$ and $\mathcal{C}'$, respectively, with $|A\symdif B|\le 2$ if and only if
    $|\mathcal{C}|+|\mathcal{C}'|-\mu(\mathcal{C},\mathcal{C}')\le k+1$.
\end{lemma}

\begin{proof}
    Suppose first that size-$k$ sets $A$ and $B$ represented by $\mathcal{C}$ and $\mathcal{C}'$, respectively, satisfy $|A\symdif B|\le 2$.
    For each $C\in\mathcal{C}$, choose $x_C\in A\cap C$, and for each $D\in\mathcal{C}'$, choose $y_D\in B\cap D$.
    Let $M_0$ be the set of the edges $\{C,D\}$ of $\confgraph{\mathcal{C}}{\mathcal{C}'}$ such that $x_C=y_D$.
    Since the members of $\mathcal{C}$ (resp.\ $\mathcal{C}'$) are pairwise disjoint, the vertices $x_C$ (resp.\ $y_D$) are pairwise distinct.
    Thus, $M_0$ is a matching in $\confgraph{\mathcal C}{\mathcal C'}$.
    Furthermore, the set of selected vertices is contained in $A\cup B$ and has size $|\mathcal{C}|+|\mathcal{C}'|-|M_0|$.
    Since $|M_0|\le\mu(\mathcal{C},\mathcal{C}')$, we obtain
    \[
    |\mathcal{C}|+|\mathcal{C}'|-\mu(\mathcal{C},\mathcal{C}') \le |\mathcal{C}|+|\mathcal{C}'|- |M_0|
    \le |A\cup B|.
    \]
    Furthermore, $|A|=|B|=k$ implies $|A\cup B|=k+|A\symdif B|/2\le k+1$.

    Conversely, let $M$ be a maximum matching in $\confgraph{\mathcal{C}}{\mathcal{C}'}$.
    Let $\mathcal{U}$ consist of the sets in $\mathcal{C}$ not saturated by $M$, and define $\mathcal{U}'$ analogously.
    Since $M$ is maximal, $C\cap D=\emptyset$ for every $C\in\mathcal{U}$ and $D\in\mathcal{U}'$.

    For each edge $e=\{C,D\}\in M$, fix $z_e\in C\cap D$.
    For each $C\in\mathcal{U}$, fix $x_C\in C$, and for each $D\in\mathcal{U}'$, fix $y_D\in D$.    
    Let
    \[
        X\coloneqq \{z_e\mid e\in M\}
        \cup\{x_C\mid C\in\mathcal{U}\}
        \cup\{y_D\mid D\in\mathcal{U}'\}.
    \]

    The sets in $\mathcal{C}$ are pairwise disjoint, and the same holds for $\mathcal{C}'$.
    Moreover, $C\cap D=\emptyset$ for every $C\in\mathcal{U}$ and $D\in\mathcal{U}'$. 
    It follows that all vertices selected in the definition of $X$ are distinct.

    Hence
    $|X|=|\mathcal{C}|+|\mathcal{C}'|-\mu(\mathcal{C},\mathcal{C}')\le k+1$.
    Moreover, $X$ intersects every set in $\mathcal{C}\cup\mathcal{C}'$.
    Since every edge of $M$ saturates one set in $\mathcal{C}$ and one set in $\mathcal{C}'$, we have $|\mathcal{U}|=|\mathcal{C}|-|M|$ and $|\mathcal{U}'|=|\mathcal{C}'|-|M|$.
    Therefore, $|X|=|M|+|\mathcal{U}|+|\mathcal{U}'| =|\mathcal{C}|+|\mathcal{C}'|-\mu(\mathcal{C},\mathcal{C}') \le k+1$.
    By its definition, $X$ intersects every set in $\mathcal{C}\cup\mathcal{C}'$.
    
    If $|X|\le k$, take an arbitrary $R\subseteq V(G)\setminus X$ with $|R|=k-|X|$ and set $A=B=X\cup R$.
    
    Suppose now that $|X|=k+1$.
    If $\mathcal{U}=\emptyset$, then $M$ saturates all sets in $\mathcal{C}$, implying $\mu(\mathcal{C},\mathcal{C}') =|M|=|\mathcal{C}|$. 
    Hence, $|X|=|\mathcal{C}|+|\mathcal{C}'|-\mu(\mathcal{C},\mathcal{C}')
    =|\mathcal{C}'|\le k$, which is a contradiction.
    Thus $\mathcal{U}\neq\emptyset$, and the same argument shows that
    $\mathcal{U}'\neq\emptyset$.

    Fix $C_0\in\mathcal{U}$ and $D_0\in\mathcal{U}'$, and define size-$k$ sets $A \coloneqq X\setminus \{y_{D_0} \}$ and $B\coloneqq X\setminus \{x_{C_0} \}$.
    Since all vertices selected in the construction of $X$ are distinct, $A$ contains $z_e$ for every $e\in M$ and $x_C$ for every $C\in\mathcal{U}$.
    Thus, $A$ intersects every set in $\mathcal{C}$, that is, $A$ is represented by $\mathcal{C}$. 
    Similarly, $B$ is represented by $\mathcal{C}'$.
    Moreover, $|A\symdif B| \le 2$ holds, completing the proof.
\end{proof}

We define the \emph{compressed configuration graph} $\mathcal{G}_{\Lclass}$ by setting $V(\mathcal{G}_{\Lclass})=\Lclass$ and joining two distinct representations $\mathcal{C},\mathcal{C}'\in\Lclass$ if $|\mathcal{C}|+|\mathcal{C}'|-\mu(\mathcal{C},\mathcal{C}')\le k+1$.
By \Cref{lem:reconf_edge_test}, an edge of $\mathcal{G}_{\Lclass}$ exists if and only if 
there exist size-$k$ sets $A,B\subseteq V(G)$ represented by $\mathcal{C}$ and $\mathcal{C}'$, respectively, that are equal or adjacent under the TJ rule.

We next describe the algorithm for an instance $(G,k,S^{\texttt{s}},S^{\texttt{t}})$.
Compute a $k$-product list $\Lclass$ and construct $\mathcal{G}_{\Lclass}$.
Let $\mathcal{S}=\{\mathcal{C}\in\Lclass\mid \mathcal{C}\text{ represents }S^{\texttt{s}}\}$ and $\mathcal{T}=\{\mathcal{C}\in\Lclass\mid \mathcal{C}\text{ represents }S^{\texttt{t}}\}$.
The algorithm returns \ttyes\ if some $\mathcal{C}^{\texttt{s}}\in\mathcal{S}$ and $\mathcal{C}^{\texttt{t}}\in\mathcal{T}$ belong to the same connected component of $\mathcal{G}_{\Lclass}$, and returns \ttno\ otherwise.

Every size-$k$ $\mathcal{F}$-deletion set $S$ is represented by some member of $\Lclass$.
Indeed, $S$ contains a minimal $\mathcal{F}$-deletion set $D$.
Since $|D|\le k$, some $\mathcal{C}\in\Lclass$ represents $D$, and then the inclusion $D\subseteq S$ implies that $\mathcal{C}$ also represents $S$.
In particular, both $\mathcal{S}$ and $\mathcal{T}$ are nonempty.

\begin{lemma}
	\label{lem:reconf_path}
    If there exist $\mathcal{C}^{\texttt{s}}\in\mathcal{S}$ and
    $\mathcal{C}^{\texttt{t}}\in\mathcal{T}$ that belong to the same
    connected component of $\mathcal{G}_{\Lclass}$, 
	then $S^{\texttt{s}}$ and $S^{\texttt{t}}$ are reconfigurable under the TJ rule.
\end{lemma}

\begin{proof}
    Let $\mathcal{C}_0,\mathcal{C}_1,\ldots,\mathcal{C}_r$ be a walk in $\mathcal{G}_{\Lclass}$ with $\mathcal{C}_0=\mathcal{C}^{\texttt{s}}$ and $\mathcal{C}_r=\mathcal{C}^{\texttt{t}}$.
    If $r=0$, the claim follows from \Cref{lem:reconf_same_node}.

    For each $i\in[r]$, \Cref{lem:reconf_edge_test} ensures that there exist size-$k$ sets $A_i$ and $B_i$ represented by $\mathcal{C}_{i-1}$ and $\mathcal{C}_i$, respectively, such that $|A_i\symdif B_i|\le 2$.
    Let $B_0=S^{\texttt{s}}$ and $A_{r+1}=S^{\texttt{t}}$.
    For every $i\in[r+1]$, both $B_{i-1}$ and $A_i$ are represented by $\mathcal{C}_{i-1}$, so \Cref{lem:reconf_same_node} implies that $B_{i-1}$ is reconfigurable to $A_i$.
    For every $i\in[r]$, $A_i$ is reconfigurable to $B_i$ in at most one TJ step.
    These sequences yield a TJ reconfiguration sequence from $S^{\texttt{s}}$ to $S^{\texttt{t}}$.
\end{proof}

\begin{lemma}
    \label{lem:reconf_reconfigurable_path}
    If $S^{\texttt{s}}$ and $S^{\texttt{t}}$ are reconfigurable under the
    TJ rule, then there exist $\mathcal{C}^{\texttt{s}}\in\mathcal{S}$ and
    $\mathcal{C}^{\texttt{t}}\in\mathcal{T}$ that belong to the same
    connected component of $\mathcal{G}_{\Lclass}$.
\end{lemma}

\begin{proof}
    Let $S_0,S_1,\ldots,S_p$ be a TJ reconfiguration sequence with $S_0=S^{\texttt{s}}$ and $S_p=S^{\texttt{t}}$.
    For each $i\in\{0,\ldots,p\}$, fix $\mathcal{C}_i\in\Lclass$ representing $S_i$.
    Such a representation exists for every set in the sequence.
    In particular, $\mathcal{C}_0\in\mathcal{S}$ and $\mathcal{C}_p\in\mathcal{T}$.

    For each $i\in[p]$, we have $|S_{i-1}\symdif S_i|=2$.
    Hence, if $\mathcal{C}_{i-1}\neq\mathcal{C}_i$, \Cref{lem:reconf_edge_test} implies that $\mathcal{C}_{i-1}$ and $\mathcal{C}_i$ are adjacent in $\mathcal{G}_{\Lclass}$.
    Removing consecutive repetitions from $\mathcal{C}_0,\mathcal{C}_1,\ldots,\mathcal{C}_p$ therefore yields a walk from $\mathcal{C}_0$ to $\mathcal{C}_p$ in $\mathcal{G}_{\Lclass}$.
    Thus, $\mathcal{C}_0$ and $\mathcal{C}_p$ belong to the same connected component of $\mathcal{G}_{\Lclass}$.
\end{proof}

Since $\Lclass$ can be computed in fixed-parameter tractable time and a maximum matching in a bipartite graph can be found in polynomial time~\cite{edmonds1965paths}, the compressed configuration graph $\mathcal{G}_{\Lclass}$ can be constructed in fixed-parameter tractable time.
This completes the proof of \Cref{thm:reconfiguration}.

As a corollary, we obtain the following.

\begin{corollary}
	\label{cor:reconfiguration}
	Both \prb{$\ell$-Forest Deletion Set Reconfiguration} and
	\prb{$r$-Pseudoforest Deletion Set Reconfiguration}
	are fixed-parameter tractable under the TJ rule when parameterized by $k + \ell$ and $k + r$, respectively.
\end{corollary}


\subsection{Connected Variant}
\label{subsec:connected}

In this subsection, we consider a connected variant of {\delprb} defined as follows.

\begin{center}
	\fbox{
		\parbox{0.9\hsize}{
			\prb{Connected {\delprb}}
			\begin{listing}{\textbf{Input:}}
				\item[\textbf{Input:}]
				An unweighted graph $G$ and a positive integer $k$.
				\item[\textbf{Task:}]
				Find an $\mathcal{F}$-deletion set $S$ of size at most $k$
				such that the induced subgraph $G[S]$ is connected.
			\end{listing}
		}
	}
\end{center}

The \prb{Connected Feedback Vertex Set} problem was introduced by Grigoriev and Sitters~\cite{connected:GrigorievS09},
who showed that the problem is NP-complete even on planar graphs and presented a polynomial-time approximation scheme (PTAS) for planar graphs.
Misra et al.~\cite{connected:MisraPRSS12} were the first to study this problem from the perspective of parameterized complexity and proved that it is fixed-parameter tractable when parameterized by $k$.
We note that their algorithm uses compact representations for feedback vertex sets introduced by Guo et al.~\cite{CompactRep:GuoGHNW06}.

In this subsection, we prove fixed-parameter tractability for the connected variant of {\delprb}.

\begin{theorem}
	\label{thm:connected}
    Let $\mathcal{F}$ be a class of graphs closed under taking subgraphs.
    Suppose that there exists an algorithm that computes a $k$-product list for $\mathcal{F}$-deletion sets in FPT time with respect to $k$.
    Then, \prb{Connected {\delprb}} is fixed-parameter tractable when parameterized by $k$.
\end{theorem}

We observe that the approach of Misra et al.~\cite{connected:MisraPRSS12} extends beyond feedback vertex sets and applies to a broad class of $\mathcal{F}$-deletion problems,
provided that a $k$-product list for $\mathcal{F}$-deletion sets can be computed in FPT time.
Their approach reduces the problem, for each $k$-product representation, to an instance of the \prb{Group Steiner Tree} problem.

\begin{center}
	\fbox{
		\parbox{0.9\hsize}{
			\prb{Group Steiner Tree (GST)}
			\begin{listing}{\textbf{Input:}}
				\item[\textbf{Input:}]
				An unweighted graph $G$,
				vertex-disjoint subsets $S_1, S_2, \ldots, S_p \subseteq V(G)$,
				and an integer $k$.
				\item[\textbf{Task:}]
				Find a tree with at most $k$ vertices that contains at least one vertex from each set $S_i$.
			\end{listing}
		}
	}
\end{center}

The sets $S_1, S_2, \ldots, S_p$ are called the \emph{Steiner groups}.
It is known that \prb{Group Steiner Tree} is fixed-parameter tractable when parameterized by $k$~\cite[Lemma~3]{connected:MisraPRSS12}.

Let $\Lclass$ be a $k$-product list for $\mathcal{F}$-deletion sets on a graph $G$.
Any connected $\mathcal{F}$-deletion set $S$ of size at most $k$ must contain a minimal $\mathcal{F}$-deletion set $S' \subseteq S$.
By definition, such a set $S'$ is represented by some $k$-product representation
$\mathcal{C} = \{C_1, C_2, \ldots, C_q\} \in \Lclass$, where $q \le k$.
Conversely, since $\mathcal{F}$ is closed under taking subgraphs, any set that contains at least one vertex from each set $C_i$ is also an $\mathcal{F}$-deletion set.

For a fixed $k$-product representation $\mathcal{C}$, deciding whether there exists a connected $\mathcal{F}$-deletion set that contains at least one vertex from each set $C_i$ reduces to an instance of \prb{Group Steiner Tree}, where the Steiner groups are $C_1, C_2, \ldots, C_q$.
Therefore, by solving the corresponding instance of \prb{Group Steiner Tree} for each $\mathcal{C} \in \Lclass$, we complete the proof of \Cref{thm:connected} and obtain the following result.

\begin{corollary}
	\label{cor:connected}
	\prb{Connected $\ell$-Forest Deletion Set} and
	\prb{Connected $r$-Pseudoforest Deletion Set}
	are fixed-parameter tractable when parameterized by $k + \ell$ and $k + r$, respectively.
\end{corollary}

\subsection{Conflict-Free Variant}
\label{subsec:conflict}

Jain et al.~\cite{conflict:JainKM20} initiated the study of the \emph{conflict-free framework} of {\delprb}.

\begin{center}
	\fbox{
		\parbox{0.9\hsize}{
			\prb{$\mathscr{F}$-Conflict-Free {\delprb}}
			\begin{listing}{\textbf{Input:}}
				\item[\textbf{Input:}]
				A graph $G$, a graph $H \in \mathscr{F}$ with $V(G)=V(H)$, and an integer $k$.
				\item[\textbf{Task:}]
				Determine whether there exists a vertex set $S \subseteq V(G)$ of size at most $k$ such that $S$ is an $\mathcal{F}$-deletion set of $G$ and an independent set of $H$.
			\end{listing}
		}
	}
\end{center}

Jain et al.~\cite{conflict:JainKM20} showed that \prb{$\mathscr{F}$-Conflict-Free {\fvsprb}} is $\W[1]$-hard parameterized by the solution size $k$ when $\mathscr{F}$ is the class of all graphs.
They also proved that the problem becomes fixed-parameter tractable when $H$ belongs to certain sparse graph classes, including $d$-degenerate graphs and nowhere dense graphs.

Subsequently, Agrawal et al.~\cite{conflict:AgrawalJKL018} strengthened these hardness results.
They proved that \prb{$\mathscr{F}$-Conflict-Free {\fvsprb}} is $\W[1]$-hard parameterized by $k$ even when $\mathscr{F}$ is restricted to graph classes of subquadratic density.
Moreover, the problem is $\W[1]$-hard even when $\mathscr{F}$ is the class of bipartite graphs.
On the positive side, Agrawal et al.~\cite{conflict:AgrawalJKL018} presented an FPT algorithm when $\mathscr{F}$ is the class of $K_{i,j}$-free graphs, for fixed integers $1 \le i \le j$, where $K_{i,j}$ denotes the complete bipartite graph with partition sizes $i$ and $j$.

Our approach can be viewed as a reinterpretation of Agrawal et al.'s algorithm~\cite{conflict:AgrawalJKL018} through the lens of product representations.
To design our FPT algorithm, we rely on the following auxiliary problem introduced in~\cite{conflict:AgrawalJKL018}.

\begin{center}
	\fbox{
		\parbox{0.9\hsize}{
			\prb{$\mathscr{F}$+Cluster Independent Set ($\mathscr{F}$+Cluster IS)}
			\begin{listing}{\textbf{Input:}}
				\item[\textbf{Input:}]
				A graph $G \in \mathscr{F}$, a cluster graph $H$ with $V(G)=V(H)$, and an integer $k$ such that $H$ has exactly $k$ connected components.
				\item[\textbf{Task:}]
				Determine whether there exists a set $S \subseteq V(G)$ of size exactly $k$ that is an independent set of both $G$ and $H$.
			\end{listing}
		}
	}
\end{center}

Agrawal et al.~\cite{conflict:AgrawalJKL018} proved that $\mathscr{F}$-CF-FVS and $\mathscr{F}$+Cluster IS are mutually reducible via FPT Turing reductions.
Consequently, fixed-parameter tractability of $\mathscr{F}$+Cluster IS implies fixed-parameter tractability of $\mathscr{F}$-CF-FVS.
We now generalize their framework to {\delprb}.

\begin{lemma}
    \label{lem:cluster_to_CF}
    Let $\mathscr{F}$ be a graph class closed under taking induced subgraphs.
    Suppose that there exists an algorithm that computes a $k$-product list for $\mathcal{F}$-deletion sets in FPT time with respect to $k$.
	If \prb{$\mathscr{F}$+Cluster IS} is fixed-parameter tractable parameterized by $k$,
	then \prb{$\mathscr{F}$-Conflict-Free {\delprb}} is fixed-parameter tractable parameterized by $k$.
\end{lemma}

\begin{proof}
	Consider an instance $(G,H,k)$ of \prb{$\mathscr{F}$-Conflict-Free {\delprb}}.

	Let $\Lclass$ be a $k$-product list for $\mathcal{F}$-deletion sets of $G$.
	For each $k$-product representation $\mathcal{C}=\{C_1,C_2, \ldots,C_q\}\in\Lclass$ with $q \leq k$, we construct an instance $(G',H',k')$ of \prb{$\mathscr{F}$+Cluster IS} as follows.

	We set $k'=q$ and define $H'$ as the cluster graph whose connected components are $C_1,C_2,\ldots,C_{k'}$.
    That is, each $C_i$ with $i \in [q]$ is a clique of $H'$.
	Let $G'$ be the subgraph of $H$ induced by $\bigcup_{i=1}^{k'} C_i$.
    Our algorithm returns {\ttyes} for $(G,H,k)$ if and only if at least one of the constructed instances $(G',H',k')$ is a $\ttyes$-instance of \prb{$\mathscr{F}$+Cluster IS}.

    Suppose first that $(G,H,k)$ is a $\ttyes$-instance of \prb{$\mathscr{F}$-Conflict-Free {\delprb}}, and let $S$ be a solution.
	That is, $S$ is an $\mathcal{F}$-deletion set of $G$ of size at most $k$ and an independent set of $H$.
	Choose a minimal $\mathcal{F}$-deletion set $D\subseteq S$ and let $\mathcal{C}\in\Lclass$ represent $D$; consider the corresponding instance $(G',H',k')$.
	By construction, $D \subseteq V(G')=V(H')$.
    Since $D$ is an independent set of $H$ and $G'$ is an induced subgraph of $H$, the set $D$ is an independent set of $G'$.
    Moreover, since $\mathcal{C}$ represents $D$, it follows that $D$ is an independent set of $H'$ and has size $k'=|\mathcal C|$.
	Hence, $(G',H',k')$ is a $\ttyes$-instance of \prb{$\mathscr{F}$+Cluster IS}.
    
    Conversely, suppose that some $(G',H',k')$ is a $\ttyes$-instance of \prb{$\mathscr{F}$+Cluster IS}, and let $S'$ be an independent set of size $k'$ in both $G'$ and $H'$.
    Let $\mathcal{C} = \{C_1, C_2, \ldots, C_q\}$ ($q \leq k$) be a $k$-product representation corresponding to $(G',H',k')$.
	Since $G'$ is an induced subgraph of $H$ and $S'$ is an independent set of $G'$, the set $S'$ is also an independent set of $H$.
	Moreover, as $S'$ contains exactly one vertex from each $C_i$ and $\mathcal{C}$ is a $k$-product representation, $S'$ is an $\mathcal{F}$-deletion set of $G$ of size at most $k$.
	Thus, $(G,H,k)$ is a $\ttyes$-instance of \prb{$\mathscr{F}$-Conflict-Free {\delprb}}.
\end{proof}

Agrawal et al.~\cite{conflict:AgrawalJKL018} focused on the class of $K_{i,j}$-free graphs and showed that \prb{$K_{i,j}$+Cluster IS} admits an FPT algorithm running in time $O^*(k^{k^2})$~\cite[Theorem~11]{conflict:AgrawalJKL018}.
Combining this fact with \Cref{lem:cluster_to_CF}, we obtain the following.

\begin{theorem}
	\label{thm:conflict}
    Let $1\leq i\leq j$ be fixed integers.
    Suppose that there exists an algorithm that computes a $k$-product list for $\mathcal{F}$-deletion sets in FPT time with respect to $k$.
	Then, \prb{$K_{i,j}$-Conflict-Free {\delprb}} is fixed-parameter tractable when parameterized by $k$.
\end{theorem}

\begin{corollary}
	\label{cor:conflict}
	\prb{$K_{i,j}$-Conflict-Free $\ell$-Forest Deletion Set} and
	\prb{$K_{i,j}$-Conflict-Free $r$-Pseudoforest Deletion Set}
	are fixed-parameter tractable when parameterized by $k + \ell$ and $k + r$, respectively.
\end{corollary}

\subsection{Simultaneous Variant}
\label{subsec:simul}

Let $\alpha$ be a positive integer.
An \emph{$\alpha$-colored graph} is a graph $G = (V, \bigcup_{i=1}^{\alpha} E_i)$ whose edge set is partitioned into $\alpha$ color classes $E_1, \ldots, E_\alpha$.
We consider a natural extension of {\delprb} to $\alpha$-colored graphs.

\begin{center}
	\fbox{
		\parbox{0.9\hsize}{
			\prb{$\alpha$-Simultaneous {\delprb}}
			\begin{listing}{}
				\item[\textbf{Input:}]
				An $\alpha$-colored graph
				$G = (V, \bigcup_{i = 1}^{\alpha} E_i)$
				and a positive integer $k$.
				\item[\textbf{Task:}]
				Find a set $S \subseteq V(G)$ of size at most $k$
				such that $S$ is an $\mathcal{F}$-deletion set of $G_i = (V, E_i)$ for every $i \in [\alpha]$.
			\end{listing}
		}
	}
\end{center}

Agrawal et al.~\cite{simultaneous:AgrawalLMS18}
showed that \prb{$\alpha$-Simultaneous Feedback Vertex Set}
can be solved in $O^*(23^{\alpha k})$ time, and admits a kernel with $O(\alpha k^{3(\alpha+1)})$ vertices.
They also proved that \prb{$\alpha$-Simultaneous Feedback Vertex Set} is $\W[1]$-hard parameterized by $k$ when $\alpha=O(\log n)$.
Thus, the dependence on the number of colors is essential when $\alpha$ is part of the input.
Furthermore, for the special case $\alpha = 2$,
they obtained an improved algorithm with running time $O^*(81^k)$.

In this subsection, our goal is to prove fixed-parameter tractability of the simultaneous variant of {\delprb}.

\begin{theorem}
	\label{thm:simul}
    Let $\mathcal{F}$ be a class of graphs closed under taking subgraphs.
    Suppose that there exists an algorithm that computes a $k$-product list for $\mathcal{F}$-deletion sets in FPT time with respect to $k$.
    Then, \prb{$\alpha$-Simultaneous {\delprb}} is fixed-parameter tractable when parameterized by $k + \alpha$.
\end{theorem}

To prove \Cref{thm:simul}, we reduce \prb{$\alpha$-Simultaneous $\mathcal{F}$-Deletion Set} to the \prb{Hitting Set} problem.
An instance of \prb{Hitting Set} consists of a hypergraph $(\mathcal{U}, \mathcal{E})$ and an integer $k$, where $\mathcal{U}$ is a finite set called the \emph{universe} and $\mathcal{E}$ is a family of subsets of $\mathcal{U}$ called \emph{hyperedges}.
The task is to determine whether there exists a subset $S \subseteq \mathcal{U}$ of size at most $k$ such that $S$ intersects every hyperedge in $\mathcal{E}$.
It is known that \prb{Hitting Set} can be solved in $O^*(2^{|\mathcal{E}|})$ time by dynamic programming
(see Section~6 in~\cite{textbook:CyganFKLMPPS15}).

Let $(G = (V, \bigcup_{i=1}^{\alpha} E_i), k)$ be an instance of \prb{$\alpha$-Simultaneous $\mathcal{F}$-Deletion Set}.
For each $i \in [\alpha]$, let $\Lclass_i$ be a $k$-product list for $\mathcal{F}$-deletion sets in $G_i$.
For each choice of $\mathcal{C}_i = \{C^i_1, C^i_2, \ldots, C^i_{q_i}\} \in \Lclass_i$ with $q_i \leq k$,
we construct an instance of \prb{Hitting Set} as follows.
The universe is $\mathcal{U} = V(G)$,
and the family of hyperedges is $\mathcal{E} = \bigcup_{i \in [\alpha]} \mathcal{C}_i$.
If there exists a tuple $(\mathcal{C}_1, \mathcal{C}_2, \ldots, \mathcal{C}_\alpha)$ such that the resulting instance $(\mathcal{U}, \mathcal{E}, k)$ of \prb{Hitting Set} is a $\ttyes$-instance,
then the algorithm returns \ttyes;
otherwise, it returns \ttno.

We now establish the correctness of the reduction.
Suppose that there exists a hitting set $S \subseteq \mathcal{U}$ of size at most $k$ for some instance $(\mathcal{U}, \mathcal{E}, k)$ constructed from a tuple $(\mathcal{C}_1, \mathcal{C}_2, \ldots, \mathcal{C}_\alpha)$.
By construction, $S$ intersects every set $C \in \mathcal{C}_i$ for each $i \in [\alpha]$.
Thus, $\mathcal{C}_i$ represents $S$ for every $i\in[\alpha]$.
Hence, since $\mathcal{F}$ is closed under taking subgraphs, $S$ is an $\mathcal{F}$-deletion set of $G_i$ for every $i \in [\alpha]$.

Conversely, suppose that $S$ is an $\mathcal{F}$-deletion set of $G_i$ of size at most $k$ for every $i \in [\alpha]$.
Then, for each $i$, the set $S$ contains a minimal $\mathcal{F}$-deletion set $S_i \subseteq S$ of $G_i$,
which is represented by some $\mathcal{C}_i \in \Lclass_i$.
Therefore, the tuple $(\mathcal{C}_1, \mathcal{C}_2, \ldots, \mathcal{C}_\alpha)$ yields a $\ttyes$-instance of \prb{Hitting Set},
and the algorithm returns \ttyes.

For each tuple, the constructed hypergraph has at most $\alpha k$ hyperedges, hence it can be solved in $O^*(2^{\alpha k})$ time.  
As a corollary, we obtain the following.

\begin{corollary}
	\label{cor:simul}
	Both
	\prb{$\alpha$-Simultaneous $\ell$-Forest Deletion Set} and
	\prb{$\alpha$-Simultaneous $r$-Pseudoforest Deletion Set}
	are fixed-parameter tractable when parameterized by
	$\alpha + k + \ell$ and $\alpha + k + r$, respectively.
\end{corollary}

\subsection{Diverse Solutions}
\label{subsec:diverse}

Let $U$ be a finite universe, and let $\mathcal{D} \subseteq 2^{U}$ denote the family of feasible solutions of a combinatorial problem.
For any two sets $A,B \subseteq U$, their \emph{symmetric difference} is defined as $A \symdif B \coloneqq (A \setminus B) \cup (B \setminus A)$.
We then consider the framework of diverse solutions on $\mathcal{D}$.

\begin{center}
	\fbox{
		\parbox{0.9\hsize}{
			\prb{Max--Min Diversification on $\mathcal{D}$}
			\begin{listing}{\textbf{Input:}}
				\item[\textbf{Input:}]
				An integer $p \geq 2$ and nonnegative integers $d$ and $k$.
				\item[\textbf{Task:}]
				Determine whether there exists a $p$-tuple $(D_1, D_2, \ldots, D_p) \in \mathcal{D}^p$ such that
				$\min_{1 \leq i < j \leq p} |D_i \symdif D_j| \geq d$ and $\max_{i \in [p]}|D_i| \leq k$.
			\end{listing}
		}}
\end{center}

\begin{center}
	\fbox{
		\parbox{0.9\hsize}{
			\prb{Max--Sum Diversification on $\mathcal{D}$}
			\begin{listing}{\textbf{Input:}}
				\item[\textbf{Input:}]
				An integer $p \geq 2$ and nonnegative integers $d$ and $k$.
				\item[\textbf{Task:}]
				Determine whether there exists a $p$-tuple $(D_1, D_2, \ldots, D_p) \in \mathcal{D}^p$ such that
				$\sum_{1 \leq i < j \leq p} |D_i \symdif D_j| \geq d$ and $\max_{i \in [p]}|D_i| \leq k$.
			\end{listing}
		}}
\end{center}

Baste et al.~\cite{diverse:BasteJMPR19} showed that both diversification problems are fixed-parameter tractable parameterized by $p+k$ when $\mathcal{D}$ is the family of all feedback vertex sets of a given graph.
Later, Kumabe~\cite{diverse:Kumabe25} obtained the same fixed-parameter tractability result for feedback vertex sets under the parameterization $p+k$, but by a different approach.
Specifically, Kumabe proposed a general framework for solving diversification problems.
This framework applies to any family of feasible solutions that admits a suitable oracle, and yields FPT algorithms for both the max--min and max--sum variants when parameterized by either $p+k$ or $p+d$, depending on the properties of the oracle.

One of the central concepts in Kumabe's framework is the notion of an \emph{exact empty extension oracle}.
Let $U$ be a finite set, $\mathcal{D} \subseteq 2^{U}$ a family of feasible solutions, $s$ a positive integer, and $Y \subseteq U$.
An \emph{exact empty extension oracle} for $\mathcal{D}$ takes $Y$ as input and either returns a set $D \in \mathcal{D}$ such that $|D| = s$ and $D \cap Y = \emptyset$, or reports that no such set exists.
Kumabe showed that if such an oracle can be implemented so that each oracle call runs in FPT time parameterized by $p+k$ and the values of $s$ and $|Y|$ are bounded by a function of $p+k$, then both \prb{Max--Min Diversification on $\mathcal{D}$} and \prb{Max--Sum Diversification on $\mathcal{D}$} are fixed-parameter tractable parameterized by $p+k$~\cite[Theorem~1]{diverse:Kumabe25}.

For the family of feedback vertex sets, Kumabe constructed such an oracle using a $k$-compact list~\cite{CompactRep:GuoGHNW06}.
We observe that the same construction applies to any $\mathcal{F}$-deletion problem, where $\mathcal{F}$ is a class of graphs closed under taking subgraphs, provided that a $k$-product list for $\mathcal{F}$-deletion sets can be computed in FPT time parameterized by $k$.

More precisely, let $G$ be a graph and let $s$ be a positive integer.
We first compute an $s$-product list $\Lclass$ for $\mathcal{F}$-deletion sets of $G$.
For each $s$-product representation $\mathcal{C} = \{C_1, C_2, \ldots, C_q\} \in \Lclass$, we check whether $C_i\setminus Y\ne\emptyset$ for every $i\in[q]$.
Whenever this condition holds, choosing one vertex from each $C_i\setminus Y$ yields a set $S$ that is disjoint from $Y$ and, by definition, is an $\mathcal{F}$-deletion set.
If such a set $S$ exists and $|S| < s$, we arbitrarily add vertices from $V(G) \setminus (S \cup Y)$ until $|S| = s$.
If fewer than $s-|S|$ such vertices are available, we proceed to the next representation.
Since $\mathcal{F}$ is closed under taking subgraphs, the resulting set $S$ remains an $\mathcal{F}$-deletion set.
Thus, since every minimal $\mathcal{F}$-deletion set of size at most $s$ is represented by some $\mathcal{C}$ in $\Lclass$, this procedure correctly implements an exact empty extension oracle.

Therefore, we obtain the following results.

\begin{theorem}
	\label{thm:diverse}
    Let $\mathcal{F}$ be a class of graphs closed under taking subgraphs.
    Suppose that there exists an algorithm that computes a $k$-product list for $\mathcal{F}$-deletion sets in FPT time with respect to $k$.
    Then, \prb{Max--Min Diversification on $\mathcal{F}$-deletion sets} and \prb{Max--Sum Diversification on $\mathcal{F}$-deletion sets} are fixed-parameter tractable when parameterized by $p + k$.
\end{theorem}

\begin{corollary}
	\label{cor:diverse_min}
    Both \prb{Max--Min Diversification on $\ell$-Forest Deletion Set} and
	\prb{Max--Min Diversification on $r$-Pseudoforest Deletion Set} are fixed-parameter tractable when parameterized by
	$p + k + \ell$ and $p + k + r$, respectively.
\end{corollary}

\begin{corollary}
	\label{cor:diverse_sum}
    Both \prb{Max--Sum Diversification on $\ell$-Forest Deletion Set} and
	\prb{Max--Sum Diversification on $r$-Pseudoforest Deletion Set} are fixed-parameter tractable when parameterized by
	$p + k + \ell$ and $p + k + r$, respectively.
\end{corollary}


\section{Concluding Remarks}
\label{sec:remarks}

In this paper, we introduced a novel data structure, called the $k$-product list, and presented FPT algorithms for its computation, focusing on two generalizations of the \textsc{Feedback Vertex Set} problem.
Our framework is versatile and yields simple and efficient FPT algorithms for a broad range of variants in this setting.

We conclude with two open problems.
First, it would be interesting to establish nontrivial lower bounds on the sizes of $k$-product lists and $k$-compact lists to better understand their combinatorial limitations.
Second, beyond feedback vertex sets and their generalizations, it remains open whether data structures that compactly encode all minimal solutions exist for other graph deletion problems, such as \prb{Odd Cycle Transversal}.
However, the $\W[1]$-hardness of \prb{Simultaneous Odd Cycle Transversal}~\cite{AgrawalKLMR17} implies that no $k$-product list for \prb{Odd Cycle Transversal} can be constructed in FPT time unless $\FPT = \W[1]$.
Therefore, an FPT-time constructible data structure representing all minimal odd cycle transversals of size at most $k$, if one exists, must use a representation other than $k$-product lists.



\newpage

\bibliography{ref}


\end{document}